\documentclass[12pt]{article}
\pdfoutput=1
\usepackage{geometry,enumerate,amsmath,amssymb}
\usepackage{fullpage}
\usepackage{graphicx}
\numberwithin{equation}{section}
\newcommand{\be}{\begin{equation}}
\newcommand{\ee}{\end{equation}}
\newcommand{\bea}{\begin{eqnarray}}
\newcommand{\eea}{\end{eqnarray}}

\renewcommand{\xi}{\gamma}
\renewcommand{\hat}{\widehat}
\renewcommand{\tilde}{\widetilde}
\renewcommand{\epsilon}{\varepsilon}
\begin{document}
\title{
\begin{flushright}\ \vskip -2cm {\small{\em DCPT-14/09}}\end{flushright}
\vskip 2cm 
Hyperbolic monopoles, JNR data and spectral curves}
\author{
Stefano Bolognesi$^{\dagger\star}$, Alex Cockburn$^\dagger$ 
and Paul Sutcliffe$^\dagger$\\[10pt]
{\em \normalsize $^\dagger$ Department of Mathematical Sciences, Durham University, Durham DH1 3LE, U.K.}\\[10pt]
{\em \normalsize 
$^\star$ Department of Physics \lq\lq E. Fermi\rq\rq\,, University of Pisa,}\\
{\em \normalsize Largo Pontecorvo, 3, Ed. C, 56127 Pisa, Italy.}\\[10pt]
{\normalsize 
s.bolognesi@durham.ac.uk, \
a.h.cockburn@durham.ac.uk, \  
p.m.sutcliffe@durham.ac.uk}
}
\date{April 2014}
\maketitle
\begin{abstract}
A large class of explicit hyperbolic monopole solutions can be obtained
from JNR instanton data, if the curvature of hyperbolic space
is suitably tuned. Here we provide explicit formulae for both the
monopole spectral curve and its rational map in terms of JNR data.
Examples with platonic symmetry are presented, together with 
some one-parameter families with cyclic and dihedral symmetries. 
These families include hyperbolic analogues of geodesics that describe symmetric
monopole scatterings in Euclidean space and we illustrate the results with
energy density isosurfaces.
There is a metric on the moduli space of hyperbolic monopoles, defined using
the abelian connection on the boundary of hyperbolic space, and we provide a
simple integral formula for this metric on the space of JNR data.
\end{abstract}

\newpage 

\section{Introduction}\quad\ 
The study of hyperbolic monopoles was initiated by Atiyah \cite{At},
using an equivalence to circle-invariant Yang-Mills instantons that
exists when there is a discrete relationship between the curvature of 
hyperbolic space and the magnitude of the Higgs field at infinity.
As in Euclidean space, twistor methods provide a correspondence between
monopole solutions and various holomorphic creatures, including 
spectral curves and rational maps. However, explicit examples of 
these holomorphic objects are rare, with even less specimens known
than in the Euclidean case. Here we remedy this situation and 
demonstrate that the hyperbolic setting is more tractable than the
Euclidean case, once the curvature of hyperbolic space is
suitably tuned.

Very recently it was shown \cite{MS} that a large class of explicit 
hyperbolic monopole solutions can be obtained from 
JNR instanton data \cite{JNR},
by restricting to the simplest example of Atiyah's discrete 
relationship, where the monopole charge is equal to the instanton number
of the associated circle-invariant instanton. 
In this paper we provide simple explicit formulae for the associated
spectral curves and rational maps directly in terms of the JNR data.
General formulae are presented and illustrated with  
examples, including some with platonic symmetry that yield new 
symmetric spectral curves. 
The simplicity and power of this approach is demonstrated via 
some one-parameter families with dihedral symmetries, 
including hyperbolic analogues of geodesics that describe symmetric
monopole scatterings in Euclidean space. 

The $L^2$ metric on the moduli space of hyperbolic monopoles is infinite, but
there is a finite metric defined using the abelian connection on the
boundary of hyperbolic space \cite{BA}. 
We present an integral formula for this metric on the space of JNR data
and confirm that it reproduces the metric of hyperbolic space for the
single monopole. Our one-parameter families of dihedral monopoles are
geodesics of this metric, because they are obtained as the fixed point 
set of a group action that is a symmetry of the metric.

\section{Hyperbolic monopoles from JNR data}\quad\ 
Hyperbolic monopoles are solutions of the Bogomolny equation
\be
D\Phi=*F,
\label{Bog}
\ee
where $F$ is the field strength of an $SU(2)$ gauge
potential $A$, and $D\Phi$ is the covariant derivative of an adjoint Higgs
field $\Phi$. 
This equation, and in particular the Hodge dual $*,$ 
is defined on hyperbolic space $\mathbb{H}^3$ of curvature
$-1,$ for which we use the
unit ball model with metric 
\be
ds^2(\mathbb{H}^3)=\frac{4(dX_1^2+dX_2^2+dX_3^2)}{(1-R^2)^2} \,,
\label{metricball0}
\ee
where $R^2=X_1^2+X_2^2+X_3^2$ and $R<1$. 

The boundary condition is that 
$|\Phi|^2=-\frac{1}{2}\mbox{Tr}\Phi^2\to v^2$ as $R\to 1,$
and the monopole charge $N\in\mathbb{Z}$ is the degree of the map
$\Phi|_{R=1}:S^2\mapsto S^2.$
If $0<2v\in\mathbb{Z}$ then hyperbolic monopoles correspond to 
circle-invariant $SU(2)$ Yang-Mills instantons in $\mathbb{R}^4$ 
with instanton number $2vN$ \cite{At}. Increasing $v$ is equivalent, after
a rescaling, to decreasing the absolute value of the curvature of
hyperbolic space, and in particular, the Euclidean limit
corresponds to $v\to\infty.$ 
In this paper we shall restrict to the simplest situation, $v=\frac{1}{2},$
where the instanton number equals the monopole charge.
In the remainder of this section we recall some results from \cite{MS} 
that apply to this tuned value.

To realise the conformal equivalence
$\mathbb{R}^4 - \mathbb{R}^2 \equiv \mathbb{H}^3 \times S^1$,
let $(x_1,x_2,x_3,x_4)$ be Cartesian coordinates in $\mathbb{R}^4$
and select the circle action that rotates the $(x_3,x_4)$ components.
The coordinates $(x_1,x_2,r),$ with $r^2=x_3^2+x_4^2,$ are then
upper half space coordinates on $\mathbb{H}^3,$ that are related to
the earlier unit ball coordinates $(X_1,X_2,X_3)$ via
\be
x_1+ix_2=\frac{2X_1+2iX_2}{1+R^2-2X_3}, \qquad
r=\frac{1-R^2}{1+R^2-2X_3}.
\ee
The $\mathbb{R}^2$ that is fixed by the circle action is the plane $r=0$,
which, after compactification, maps in the ball model to the boundary 
$S^2$ given by $R=1.$

The JNR ansatz \cite{JNR} is an extension of the 't Hooft ansatz \cite{tH}
and provides a construction of some 
$N$-instantons in terms of a harmonic function in $\mathbb{R}^4$,
specified by the location of $N+1$ poles and associated 
positive real weights. The circle invariance of the instanton is
obtained by placing all the poles in the plane $r=0,$
to give the harmonic function
\be \psi=\sum_{j=0}^N \frac{\lambda_j^2}{|x_1+ix_2-\xi_j|^2+r^2}, \label{JNR}\ee
specified by the complex constants $\xi_j, \ j=0,\ldots N,$ 
together with their positive real weights $\lambda_j^2.$ 
The formulae in \cite{MS} provide expressions for the gauge potential
and the Higgs field of the hyperbolic monopole in terms of this harmonic 
function. In particular, these expressions can be used to obtain the
following formula for the squared magnitude of the Higgs field 
in upper half space coordinates
\be
|\Phi|^2=\frac{r^2}{4\psi^2}\bigg(
\bigg(\frac{\partial\psi}{\partial x_1}\bigg)^2+
\bigg(\frac{\partial\psi}{\partial x_2}\bigg)^2
+\bigg(\frac{\psi}{r}+\frac{\partial\psi}{\partial r}\bigg)^2\bigg).
\label{higgs}
\ee
The monopole energy density is obtained by applying
the Laplace-Beltrami operator to $|\Phi|^2.$  

The symmetries of a hyperbolic monopole are most readily seen in the
ball model, where the poles $\gamma_j$ correspond to the
Riemann sphere coordinates of $N+1$ points on the $S^2$ boundary 
of $\mathbb{H}^3.$ 
If the weights are chosen to be 
\be
\lambda_j^2=1+|\xi_j|^2,
\label{can}
\ee
then they are all equal after a conformal transformation
to the unit ball model. We shall refer to the choice (\ref{can}) as 
canonical weights. For canonical weights the 
symmetry of the set of points $\{\xi_j\}$,
regarded as points on the Riemann sphere,
is inherited as a symmetry of the hyperbolic monopole. 
This Riemann sphere is the boundary of the unit ball model and 
spatial rotations act as $SU(2)$ M\"obius transformations on 
the Riemann sphere.

The JNR ansatz (\ref{JNR}) reduces to the 't Hooft ansatz \cite{tH}
\be \psi=1+\sum_{j=1}^N \frac{\lambda_j^2}{|x_1+ix_2-\xi_j|^2+r^2}
\label{tHooft} \ee
by taking the limit $\lambda_0^2=1+|\xi_0|^2\to\infty.$
Thus in considering the symmetry of a monopole obtained from the
't Hooft form one must bear in mind that there is a pole,
with canonical weight, at the
point $\infty$ on the Riemann sphere.

The $N$-monopole moduli space $\mathbb{M}_N$ has dimension $4N-1$ 
but the JNR ansatz (\ref{JNR}) has $3N+3\,$ real parameters.
The associated monopole fields are unchanged if $\psi$ is multiplied by 
a constant, so only the relative weights are of relevance in the JNR form,
which reduces the JNR parameter count by one to $3N+2.$
If $N\ge 3$ then all these parameters are independent and the JNR
construction produces a $(3N+2)$-dimensional subspace 
${\mathbb{M}}^{\mbox{\tiny{JNR}}}_N$
of the $(4N-1)$-dimensional monopole moduli space $\mathbb{M}_N$. 
Note that for $N=3$ this
construction gives the full 11-dimensional moduli space and
${\mathbb{M}}^{\mbox{\tiny{JNR}}}_3={\mathbb{M}_3}.$
If $N=1$ then the JNR ansatz is equivalent to the 't Hooft ansatz
(\ref{tHooft}), as the two can be related by an action of the conformal
group. This reduces the parameter count to 3, which is the
correct dimension and 
${\mathbb{M}}^{\mbox{\tiny{JNR}}}_1={\mathbb{M}_1}.$
For $N=2$ there are three poles, which therefore automatically lie on 
a circle. For poles on a circle there is an action of the
conformal group that moves the poles around the circle and acts on their
weights \cite{JNR}. This reduces the number of independent parameters in the JNR
ansatz by one, leaving the correct dimension 7 
and ${\mathbb{M}}^{\mbox{\tiny{JNR}}}_2={\mathbb{M}_2}.$

In summary, for the value $v=\frac{1}{2}$ in hyperbolic space of 
curvature $-1,$ we have the result that
for $N\le 3$ 
the dimension of the moduli space of JNR generated $N$-monopoles is
$\mbox{dim}({\mathbb{M}}^{\mbox{\tiny{JNR}}}_N)=\mbox{dim}({\mathbb{M}}_N)
=4N-1,$
and all monopoles can be obtained using the JNR construction.
However, if $N>3$ then
$\mbox{dim}({\mathbb{M}}^{\mbox{\tiny{JNR}}}_N)
=3N+2<4N-1=\mbox{dim}({\mathbb{M}}_N),$
so a large class of monopoles can be obtained using the JNR construction,  
rather than all monopoles. Note that any monopole obtained from the JNR data 
(\ref{JNR})
can be acted upon by a spatial rotation to map it to a monopole that is
obtained from the 't Hooft data (\ref{tHooft}). The required spatial rotation
is simply one that rotates any of the $N+1$ poles on the Riemann sphere 
to the point $\infty.$ The fact that this pole has canonical weight in
't Hooft form is no loss of generality, because in JNR form only the 
relative weights are relevant.

\section{Spectral curves and rational maps from JNR data}\quad\
The theory of spectral curves for hyperbolic monopoles 
was introduced by Atiyah \cite{At} and closely 
parallels the Euclidean case pioneered by Hitchin \cite{Hi1}.
The salient features of relevance to the current paper are
briefly reviewed below, but for more details consult the papers 
\cite{MuSi1,MuSi2,MNS,NoRo}. 

The spectral curve of a charge $N$ hyperbolic monopole 
is a biholomorphic curve, 
of bidegree $(N,N),$ in 
$\mathbb{CP}^1 \times\mathbb{CP}^1.$ 
 Let $(\eta,\zeta)\in \mathbb{CP}^1 \times\mathbb{CP}^1$ represent two
points on the Riemann sphere, regarded as the boundary of the unit ball
model of hyperbolic space. We associate to the pair $(\eta,\zeta)$ the
oriented geodesic from $\hat\eta=-1/\bar\eta$ (the antipodal point to $\eta$)
to $\zeta$. As the start and end points of the geodesic  
cannot coincide then the anti-diagonal $\hat\eta=\zeta$ must be removed. 
The pair $(\eta,\zeta)$ is then a point in the space of oriented geodesics,
which is the twistor space of $\mathbb{H}^3.$

The spectral curve is a complex curve in twistor space 
that corresponds to the set of all geodesics along which the equation
\be
(D_s-i\Phi)w=0
\label{hitchin}
\ee
has a normalisable solution, where $s$ is arc length along the geodesic
and $w$ is a complex two-component scalar. 
The spectral curve takes the form
\be \sum_{i=0,j=0}^N c_{ij}\eta^i\zeta^j=0, 
\label{gensc}
\ee
where $c_{ij}$ are complex constants.
There is a reality condition on the spectral curve, that derives from
reversing the orientation of the geodesic, and this produces the
conditions
\be
\bar c_{ij}=(-1)^{N+i+j}c_{N-j,N-i}.
\label{reality}
\ee
In terms of ball coordinates, a single monopole with position $(X_1,X_2,X_3)$
has the spectral curve
\be
2\eta\zeta(X_1-iX_2)+\zeta(1+R^2-2X_3)-\eta(1+R^2+2X_3)-2(X_1+iX_2)=0.
\label{star}
\ee
This spectral curve is known as a star and corresponds to the set of 
all geodesics through the point $(X_1,X_2,X_3)$. 
The star is more usually presented in terms of upper half space
coordinates $(x_1,x_2,r)$ for the position of the monopole, which yields
\be
\eta\zeta (x_1-ix_2)+\zeta-\eta(x_1^2+x_2^2+r^2)-(x_1+ix_2)=0.
\label{staruhs}
\ee
A curve of the form (\ref{gensc}), 
with coefficients satisfying (\ref{reality}), must obey certain
constraints \cite{At,MuSi1,MNS} to be a spectral curve. These constraints
can be written in terms of conditions for standard line bundles defined
on the spectral curve and can be translated into the
language of function theory on the spectral curve, which is a
Riemann surface of genus $(N-1)^2.$ In this setting the conditions become
relations that must be satisfied by integrals of certain holomorphic
differentials around particular cycles of the Riemann surface. 
This is a difficult problem in algebraic geometry and the general
solution is not available beyond the elliptic case.
It is therefore a highly non-trivial task to explicitly enforce these 
conditions for multi-monopoles and
the only known examples are for 2-monopoles 
(when the spectral curve is elliptic), 
a tetrahedral 3-monopole and a cubic 4-monopole.
The two platonic examples are tractable, even though the genus of the 
curve is greater than one, because in both cases the curve is the 
Galois cover of an elliptic curve: although even with this 
significant simplification the calculation is 
still quite involved \cite{NoRo}.  

For the tuned case of $v=\frac{1}{2},$
we now describe how to sidestep the algebraic geometry and obtain an
explicit expression for the spectral curve in terms of the free JNR data of
poles and weights. The starting point is the identification, mentioned earlier,
 of a hyperbolic $N$-monopole with $2v\in\mathbb{Z}$ and a
circle-invariant instanton with 
instanton number $I=2vN.$ The ADHM construction \cite{ADHM} provides a
transformation between instantons and 
quaternionic $(I+1)\times I$ matrices
${\cal M}$ satisfying the ADHM equation, which is the 
reality condition that ${\cal M}^\dagger{\cal M}$
is a real matrix, where $^\dagger$ denotes the quaternionic conjugate 
transpose.  

Braam and Austin \cite{BA} made a detailed investigation of the imposition
of circle invariance on the ADHM equation and found that this results in a
nonlinear difference equation for complex $N\times N$ matrices defined on
a one-dimensional lattice with $2v$ lattice sites, plus appropriate boundary
conditions at the end points of the lattice. The $2v$ lattice sites appear
as labels for the weights under the circle action of blocks of the ADHM
matrix ${\cal M}.$
This difference equation is a discrete Nahm equation, in that in
the continuum limit $v\to\infty$ it becomes the Nahm equation, which is
an ordinary differential equation for complex $N\times N$ matrices with
solutions that map to Euclidean $N$-monopoles under the Nahm transform 
\cite{Nahm}. This limit to the Nahm equation for Euclidean monopoles 
is quite natural, in that the continuum limit $v\to\infty$ is equivalent
to the limit in which the curvature of hyperbolic space tends to zero.

The discrete Nahm equation is an integrable system with an 
associated spectral curve that encodes the conserved quantities of the
discrete evolution along the lattice.
 Furthermore, this spectral curve is indeed the
spectral curve of the hyperbolic monopole \cite{MuSi2}.
This parallels a similar result in Euclidean space, 
identifying the spectral curve of the
integrable Nahm equation with the spectral curve of the Euclidean monopole
\cite{HiMu}. The formulae in \cite{MuSi2} provide an equation for the
spectral curve in terms of the matrices that solve the discrete Nahm
equation, so in principle this provides an alternative approach to 
calculating the spectral curve. However, until now, 
this approach has not been exploited, except for the case of the single 
monopole and the axial 2-monopole, because of the apparent 
difficulty in solving the discrete Nahm equation. 

Let us now restrict to the situation of interest in the current paper,
namely $v=\frac{1}{2}.$ At first glance it may appear that the discrete 
Nahm equation with only one lattice site is too degenerate to contain 
any useful information. Indeed it is true that in this anti-continuum limit
there is no discrete evolution, so there is no discrete equation to solve.
However, the boundary condition does survive and so does the spectral curve,
which is obtained by evaluation on the single lattice site.
The boundary condition simply becomes the original ADHM equation
but restricted to complex rather than quaternionic matrices. This is to be
expected as $v=\frac{1}{2}$ corresponds to the appearance in the ADHM matrix
of only the trivial weight under the circle action. 

Explicitly, if we write the complex ADHM matrix in standard form 
it is given by
\be
{\cal M}=\begin{pmatrix} L \\ M \end{pmatrix},
\label{ADHMmatrix}
\ee
where $M$ is a complex $N\times N$ symmetric matrix and $L$ is an
$N$-component complex vector. This is required to satisfy the ADHM equation,
$\Im({\cal M}^\dagger{\cal M})=0$, where $^\dagger$ is now simply the
complex conjugate transpose and $\Im$ denotes the imaginary part.
In terms of this notation, the spectral curve formula in \cite{MuSi2}
evaluated on the single lattice site simplifies to
\be
\mbox{det}\big(\eta\zeta M^\dagger+\zeta-\eta {\cal M}^\dagger {\cal M}-M\big)=0.
\label{det}
\ee
The 't Hooft form of the instanton (\ref{tHooft}) corresponds to the
ADHM matrix
\be
{\cal M}=\begin{pmatrix}
\lambda_1 & \lambda_2  & \cdots & \lambda_N \\
\xi_1 & & & \\
&  \xi_2 & & \\
& & \ddots & \\
& & &  \xi_N
\end{pmatrix}.
\label{tHooftM}
\ee
Extending this to the more general JNR form (\ref{JNR}) is a little
more involved because the JNR data does not come in a natural format
to fit into the standard form of the ADHM matrix, so an
appropriate change of basis needs to be found. 
Explicitly, the ADHM matrix is given in terms of the JNR data by
\be
{\cal M}=S\Gamma V, \quad
\mbox{where} \quad
\Gamma=
\begin{pmatrix}
\lambda_1 \xi_0 & \lambda_2 \xi_0 & \cdots & \lambda_N \xi_0\\
\lambda_0 \xi_1 & & & \\
& \lambda_0 \xi_2 & & \\
& & \ddots & \\
& & & \lambda_0 \xi_N
\end{pmatrix}.
\label{JNRM}
\ee
Here $S\in O(N+1)$ and $V\in GL(N,\mathbb{R})$ 
perform the required change of basis and must
satisfy the equation
\be 
S
\begin{pmatrix}
\lambda_1  &  \lambda_2 & \cdots & \lambda_N \\
\lambda_0  & & & \\
& \lambda_0  & & \\
& & \ddots & \\
& & & \lambda_0 
\end{pmatrix}
V
= \begin{pmatrix}
0  &  0 & \cdots & 0 \\
1  & & & \\
& 1  & & \\
& & \ddots & \\
& & & 1
\end{pmatrix}.
\ee
For $N=1$ and $N=2$ the required matrices $S$ and $V$ can be found
in \cite{Os} and in the special case that all the $N+1$ weights are equal the
matrices are presented in \cite{AS} for arbitrary $N.$ 
Here we require the general solution, which we find to be  
\be
V_{ij}=\begin{cases}
0 & \text{if\ } i>j \\
{p_i}/({\lambda_0p_{i-1}}) & \text{if\ } i=j \\
-\lambda_i\lambda_j p_jp_{j-1}/\lambda_0 & \text{if\ } i<j \\
\end{cases}
\label{vmatrix}
\ee
and 
\bea
S_{i1}&=& \lambda_0\lambda_{i-1}p_{i-1}p_{i-2}  
\ \ \quad \text{\ for \ } i=1,\ldots,N+1 \nonumber \\
S_{1j}&=& -\lambda_{j-1}p_N  
\qquad\qquad \text{\ for \ } j=2,\ldots,N+1 \label{smatrix} \\
S_{ij}&=& \lambda_0 V_{j-1,i-1}  
\qquad\quad\ \,\text{\ for \ } i,j=2,\ldots,N+1, \nonumber
\eea
where we have introduced the notation
$p_i=(\sum_{j=0}^i\lambda_j^2)^{-1/2},$ for $i=0,\ldots,N$ \ 
together with $p_{-1}=p_N$ and $\lambda_{-1}=\lambda_0.$
It can be checked that this general solution reduces to the previously
known special cases in \cite{Os,AS}.
The 't Hooft case is recovered in the limit 
$\lambda_0^2=1+|\gamma_0|^2\to\infty,$ 
where $S$ and $\lambda_0V$ both become the identity matrix.

Substituting the above expressions into the formula (\ref{det}) provides
an explicit construction of the spectral curve in terms of JNR data.
Although this appears to be a rather cumbersome procedure, in fact it yields
a very elegant result, as we now explain. First of all, for 't Hooft
data the diagonal form of $M$ in the ADHM matrix (\ref{tHooftM}) allows the 
determinant formula (\ref{det}) to be easily calculated, producing the result
 \be
\prod_{j=1}^N(\zeta-\gamma_j)(1+\eta\bar\gamma_j)
-\eta\sum_{j=1}^N\lambda_j^2\mathop{\prod_{k=1}^N}_{k\ne j}
(\zeta-\gamma_k)(1+\eta\bar\gamma_k)=0.
\label{thooftspectralcurve}
\ee
Note that, as required, this formula is invariant under permutations of
the $N$ poles, $\gamma_j$ for $j=1,\ldots,N$ together with their weights
$\lambda_j^2.$ Next we recall that 't Hooft data is simply JNR data with
a pole at $\infty$ with canonical weight. Furthermore, we know how the poles and
weights transform under a rotation, given by an $SU(2)$ M\"obius 
transformation. By applying this transformation to the spectral curve
(\ref{thooftspectralcurve}) we obtain the following elegant formula for the 
spectral curve in terms of JNR data 
\be
\sum_{j=0}^N\lambda_j^2\mathop{\prod_{k=0}^N}_{k\ne j}
(\zeta-\gamma_k)(1+\eta\bar\gamma_k)=0.
\label{JNRspectralcurve}
\ee
Equation (\ref{JNRspectralcurve}) is one of the main results of this paper,
providing a simple explicit formula for the spectral curve in terms of 
free JNR data. There is an obvious invariance of this formula under 
permutations of all $N+1$ poles, together with their weights, and 
it degenerates to the formula (\ref{thooftspectralcurve}) in the 't Hooft limit 
$\lambda_0^2=1+|\gamma_0|^2\to\infty.$ 
An obvious consequence of equation (\ref{JNRspectralcurve}) is
that the spectral curve contains all geodesics that connect any pair of 
JNR poles.  
Before we go on to present some example spectral curves using this formula,
we shall first consider the construction of rational maps from JNR data.

Atiyah \cite{At} introduced a correspondence between hyperbolic $N$-monopoles
and degree $N$ based rational maps between Riemann spheres, modulo 
multiplication by a constant phase. We denote the rational map by
${\cal R}(z)$ and the based condition is that ${\cal R}(\infty)=0,$ so that
the map is a ratio of two polynomials where the denominator has degree $N$
and the numerator has degree less than $N.$ 
It describes the scattering data of equation (\ref{hitchin}) along 
geodesics that start at $\hat\eta=\infty$ and end at $\zeta=z.$
In more detail, one considers the solution of equation (\ref{hitchin})
that decays at the $\hat\eta=\infty$ end and defines ${\cal R}(z)$ to be
the ratio of the decaying to the growing component at the $\zeta=z$ end.
This implies that the denominator of the rational map is the spectral curve 
after the substitution $(\eta,\zeta)=(0,z),$ 
since the spectral curve specifies
 the geodesics along which there is no growing component at either end. 

Following Donaldson's derivation \cite{Do} of the rational map for
a Euclidean monopole from the solution of the Nahm equation, Braam and 
Austin \cite{BA} 
obtained a similar formula for the rational map of a hyperbolic
monopole from the solution of the discrete Nahm equation.
Restricting their formula to the $v=\frac{1}{2}$ case, and using our notation
(\ref{ADHMmatrix}) for the ADHM matrix, this becomes 
\be {\cal R}(z)=L(z-M)^{-1}L^t. 
\label{ratmap}
\ee

\noindent The rational map takes a particularly simple form for 't Hooft data,
because $M$ is diagonal in the ADHM matrix (\ref{tHooftM}).
Applying (\ref{ratmap}) in this case yields
\be
{\cal R}=\sum_{j=1}^N \frac{\lambda_j^2}{z-\xi_j},
\label{tHooftR}
\ee
which reveals that the interpretation of the 't Hooft parameters as poles
and weights in the harmonic function that determines the instanton 
conveniently extends to the same interpretation of poles and weights for 
the rational map.

The generalization of the rational map formula (\ref{tHooftR}) 
to the JNR case is more complicated. In particular, it cannot be 
obtained using the same rotation trick that we used to obtain the 
JNR spectral curve from the 't Hooft case, because the rational map involves
scattering along geodesics that originate at $\infty$ and only rotations
that leave this point fixed can be applied. We therefore require the 
following alternative strategy to determine the JNR rational map.
The denominator is obtained by using the
fact that it is equal (up to a constant multiple) to the  
spectral curve (\ref{JNRspectralcurve}) evaluated at
$(\eta,\zeta)=(0,z).$ The numerator is then obtained by the requirement
that the rational map must be invariant 
under any permutation of the $N+1$ poles and weights, together with the
fact that it must reduce to the 't Hooft rational map 
(\ref{tHooftR}) in the limit  $\lambda_0^2=1+|\gamma_0|^2\to\infty.$
The final result is 
\be
{\cal R}=\bigg\{
\sum_{i=0}^N\sum_{j=i+1}^N\lambda_i^2\lambda_j^2(\gamma_i-\gamma_j)^2
\mathop{\prod_{k=0}^N}_{k\ne i,j}(z-\gamma_k)\bigg\} / 
\bigg\{
\bigg(\sum_{i=0}^N\lambda_i^2\bigg)
\bigg(\sum_{j=0}^N\lambda_j^2\mathop{\prod_{k=0}^N}_{k\ne j}(z-\gamma_k)\bigg)
\bigg\}.
\label{JNRratmap}
\ee
In the appendix we prove this formula directly using the definition
(\ref{ratmap}) together with the ADHM matrix (\ref{JNRM}) and the change of
basis matrices (\ref{vmatrix}) and (\ref{smatrix}).

In the following section we illustrate the use of our spectral curve and
rational map formulae by calculating some examples with platonic symmetry.
However, we first conclude this section by considering the single monopole 
and the axial $N$-monopole. 

For $N=1$ the 't Hooft form gives all 1-monopoles and the spectral curve
is the star (\ref{staruhs}) with point $x_1+ix_2=\xi_1$ and $r=\lambda_1.$  
In particular, taking $\xi_1=0$ with canonical weight gives the spectral
curve $\eta-\zeta=0,$ for a 1-monopole at the origin, with
rational map ${\cal R}=1/z.$

Taking canonical weights and 
$\xi_j=\omega^j,$ for $j=0,\ldots,N,$ where $\omega=e^{\frac{2\pi i}{N+1}},$ 
yields the axially symmetric spectral curve
\be
\sum_{i=0}^N (-1)^i\eta^i\zeta^{N-i}=0
\label{axial}
\ee
 and the rational map
${\cal R}={1}/{z^N}.$
Although the set of poles appears to have only a dihedral $D_{N+1}$ symmetry,
the enhancement to axial symmetry is a consequence of the previously mentioned
fact that when all poles lie on a circle there is an action of the 
conformal group that moves the poles around the circle and acts on 
their weights. The axial symmetry is manifest in the spectral curve
(\ref{axial}) as the invariance under 
$(\eta,\zeta)\to(e^{i\theta}\eta,e^{i\theta}\zeta)$, corresponding to
 a rotation around the $X_3$-axis by an arbitrary angle $\theta.$
The symmetry is evident in the rational map as the relation 
${\cal R}(e^{i\theta}z)=e^{-iN\theta}{\cal R}(z),$ where we recall that a 
rational map is defined modulo multiplication by a constant phase.

This is one of the few examples in which the full symmetry of the monopole
is apparent from the rational map, because the action of this symmetry group
happens to fix the point $z=\infty.$ If a monopole is symmetric under a
transformation that moves the point $\infty$ on the Riemann sphere 
boundary of hyperbolic space, then the rational map cannot detect
this symmetry, because in general 
it is not known how to explicitly relate the based 
rational map ${\cal R}(z)$, with ${\cal R}(\infty)=0$, to a
rational map that is based at a different point than $\infty.$  

Note that if the monopole is of JNR type then the formula 
(\ref{JNRratmap}) allows the calculation of the rational map based at
an arbitrary point $z_\infty$, since we know how the M\"obius 
transformation that moves $z_\infty$ to $\infty$ acts on the JNR poles and
weights. We can then apply (\ref{JNRratmap}) to the rotated poles and
weights and finally obtain the rational map based at $z_\infty$ by 
replacing $z$ by its image under the M\"obius transformation. 

The above axial monopoles are positioned at the point $(X_1,X_2,X_3)=(0,0,0),$
but for future reference it will be useful to have the
spectral curve of the axial 2-monopole with position 
$(X_1,X_2,X_3)=(0,0,b).$  This is obtained by taking canonical weights 
with poles
$\xi_j={\frac{(1+b)}{(1-b)}}\omega^j$ for $j=0,1,2,$ 
where $\omega=e^{{2\pi i}/{3}}$.
The resulting spectral curve is
\be
(1+b)^4\eta^2+(1-b)^4\zeta^2-{(1-b^2)^2}\eta\zeta=0.
\label{axial2shifted}
\ee

\section{Platonic spectral curves}\quad\
Platonic $N$-monopoles can be obtained by taking canonical weights with 
the JNR poles as the roots of a Klein vertex polynomial \cite{Kle} for
a platonic solid with $N+1$ vertices. 
Examples of explicit Higgs fields and monopole energy densities
for platonic monopoles were presented in \cite{MS}, 
essentially using the formula
(\ref{higgs}) for the squared magnitude of the Higgs field $|\Phi|^2$, 
together with the fact that the monopole energy density is obtained by applying
the Laplace-Beltrami operator to $|\Phi|^2.$ However, spectral curves were
not discussed in that paper, so the results in this section are complementary
to that study. Furthermore, although 
some expressions for rational maps appear in \cite{MS}, it
is important to recognize that those rational maps are 
compatible with the $SO(3)$ action on the hyperbolic ball,
being a hyperbolic analogue of Jarvis rational maps \cite{Jar}
defined for Euclidean monopoles, rather
than the hyperbolic analogue of Donaldson rational maps \cite{Do} 
discussed in the current paper.

The lowest charge example of a platonic monopole is the tetrahedral 3-monopole,
obtained by taking the roots of 
the Klein polynomial associated with the vertices of a tetrahedron
\be  
{\cal T}_v(\xi)=\xi^4+2i\sqrt{3}\xi^2+1.
\ee
Explicitly, the poles are
$\xi_0=\frac{1+i}{\sqrt{3}+1}, \
\xi_1=-\xi_0, \ \xi_2=\xi_0^{-1}, \ \xi_3=-\xi_0^{-1},$  
and equation (\ref{JNRspectralcurve}) 
with canonical weights
gives the spectral curve
\be
(\eta-\zeta)^3+\frac{i}{\sqrt{3}}(\eta+\zeta)(\eta\zeta+1)(\eta\zeta-1)=0.
\label{tet3}
\ee
This spectral curve was derived previously in \cite{NoRo},
using the methods of algebraic geometry that we mentioned earlier.
This curve is invariant under the
generators of the tetrahedral group
\be
(\eta,\zeta)\mapsto(-\eta,-\zeta),\qquad \qquad
(\eta,\zeta)\mapsto\bigg(\frac{\eta-i}{\eta+i},\frac{\zeta-i}{\zeta+i}\bigg).
\label{tetrotations}
\ee
Note that restricting the curve (\ref{tet3}) 
to the diagonal $\eta=\zeta$ determines the spectral geodesics that pass
through the origin as
\be
{\cal T}_{e}(\zeta)=\zeta(\zeta^4-1)=0,
\label{kleintete}
\ee
where we recognize ${\cal T}_e$ 
as the Klein polynomial 
for the edges of the tetrahedron.
Applying formula (\ref{JNRratmap}) allows us to obtain the associated
rational map 
\be {\cal R}=\frac{5iz^2+\sqrt{3}}{\sqrt{3}z^3+iz}, \ee
where the $C_2$ symmetry is manifest, ${\cal R}(-z)=-{\cal R}(z)$, but
not the full tetrahedral symmetry.

The octahedral 5-monopole is obtained from six poles 
(with canonical weights) on the vertices of an octahedron, given by 
the roots of the Klein polynomial 
\be
{\cal O}_v(\xi)=\xi(\xi^4-1),
\ee
including the root at $\infty.$
As one pole is at $\infty$ this example is of 't Hooft form and 
applying formula (\ref{thooftspectralcurve}) results in the spectral curve
\be
(\eta-\zeta)\bigg(
(\eta^4-1)(\zeta^4-1)+8\eta\zeta(\eta^2+\zeta^2)
\bigg)=0,
\label{oct5}
\ee
which is invariant under the generators of the octahedral group 
\be
(\eta,\zeta)\mapsto(i\eta,i\zeta),\qquad
(\eta,\zeta)\mapsto\bigg(\frac{\eta-i}{\eta+i},\frac{\zeta-i}{\zeta+i}\bigg).
\label{octrotations}
\ee
Restricting to the diagonal $\eta=\zeta$ makes the first factor in
(\ref{oct5}) vanish identically but the condition that the second factor
also vanishes is 
\be
{\cal O}_{f}(\zeta)= \zeta^8+14\zeta^4+1=0,
\label{kleinoctf}
\ee
where ${\cal O}_f$ is the  
Klein polynomial for the face centres of the octahedron.
Equation (\ref{tHooftR}) for the rational map from 't Hooft data yields
\be
{\cal R}=\frac{9z^4-1}{z^5-z},
\ee
with denominator equal to the Klein polynomial ${\cal O}_v(z).$
Note that the fact that the denominator of the rational map is the
Klein polynomial for the vertices of the polyhedron
 is generic if the Klein polynomial is in an 
orientation in which there is a root at $\infty.$ 
This follows immediately from (\ref{tHooftR}).  

As a final platonic example of the ease of generating spectral curves using our
approach, we consider the icosahedral 11-monopole. 
The vertex Klein polynomial for the icosahedron is
\be
{\cal Y}_v(\xi)=\xi^{11}+11\xi^6-\xi,
\label{kleinyv}
\ee
where the orientation is such that one root is at $\infty.$
Taking the canonical weight poles as the roots of (\ref{kleinyv}) 
and using (\ref{thooftspectralcurve}) we obtain the substantial
spectral curve
\bea
&&(\eta-\zeta)\bigg(
\eta^{10}\zeta^{10}
+11(\eta^{10}\zeta^{5}+\eta^{5}\zeta^{10}-\eta^{5}-\zeta^{5})
-75(\eta^{9}\zeta^{6}+\eta^{6}\zeta^{9}-\eta^{4}\zeta-\eta\zeta^{4})
\nonumber \\ &&
-50(\eta^{8}\zeta^{7}+\eta^{7}\zeta^{8}-\eta^{3}\zeta^{2}-\eta^{2}\zeta^{3})
+25(\eta^{9}\zeta+\eta\zeta^{9}-\eta^{8}\zeta^{2}-\eta^{2}\zeta^{8})
+100(\eta^{7}\zeta^{3}+\eta^{3}\zeta^{7})
\nonumber \\ &&
-225(\eta^{6}\zeta^{4}+\eta^{4}\zeta^{6})
+746\eta^{5}\zeta^{5}-\eta^{10}-\zeta^{10}
+1
\bigg)=0,
\label{icos11}
\eea
that is invariant under the following generators of the icosahedral group,
where $\omega=e^{2\pi i/5}$,
\bea
(\eta,\zeta)&\mapsto& (\omega\eta,\omega\zeta),\\
(\eta,\zeta)&\mapsto& \bigg(
\frac{(\omega^3-1)\eta+\omega-\omega^2}{(\omega-\omega^2)\eta+1-\omega^3},
\frac{(\omega^3-1)\zeta+\omega-\omega^2}{(\omega-\omega^2)\zeta+1-\omega^3}
\bigg).
\eea
The first factor in (\ref{icos11}) automatically vanishes on the
diagonal $\eta=\zeta$ and the second factor vanishes when
\be
{\cal Y}_f(\zeta)=\zeta^{20}-228\zeta^{15} 
+494\zeta^{10} +228\zeta^{5}+1=0,
\ee
which is the Klein polynomial for the face centres of the icosahedron.
For this example the rational map is
\be
{\cal R}=\frac{26z^{10}+86z^5-1}{z^{11}+11z^6-z},
\ee
where the denominator is indeed ${\cal Y}_v(z)$, with
the obvious $C_5$ symmetry ${\cal R}(\omega z)={\cal R}(z)/\omega.$

\section{Dihedral one-parameter families}\quad\ 
In Euclidean space the geodesic approximation \cite{Ma1} can be used to
interpret particular one-parameter families of static monopoles in terms 
of monopole dynamics. In hyperbolic space this interpretation is not so clear,
and we shall discuss this aspect further in section \ref{sec-metric}. 
Therefore, for now, one should simply regard the results in this section as
some interesting one-parameter families of symmetric static hyperbolic
 monopoles. 
However, as we shall see, they bear a striking resemblance to similar 
symmetric families in Euclidean space that indeed describe symmetric
monopole scattering. Regarding these results as hyperbolic analogues
of Euclidean monopole scattering is a reasonable point of view. 

Our strategy is to impose a symmetry on the hyperbolic $N$-monopole 
that is an appropriate finite subgroup of the $SO(3)$ rotational symmetry
group, so that the resulting fixed point set is a one-parameter family of 
monopoles. Dihedral symmetry is particularly fruitful in this context and
is a natural extension of the results in the previous section, since the 
platonic symmetry groups have dihedral subgroups. Of course, we shall actually
be imposing the symmetry within the moduli space 
${\mathbb{M}}^{\mbox{\tiny{JNR}}}_N$, and there are three different ways to
obtain symmetric families of JNR data, as follows. The first type of 
one-parameter family involves moving the positions of the poles around the
Riemann sphere with the associated weights at their canonical values. 
The second type involves fixed positions for the poles but a 
variation of the weights from their canonical values. Finally, the third
type involves simultaneously varying the positions of the poles 
together with non-canonical weights. We shall provide examples of all
three types of families with dihedral symmetry. Dihedral symmetry is not
the only finite symmetry group that is useful in generating families of
monopoles, as we illustrate with a cyclic and a tetrahedral example.

\begin{figure}
\begin{center}
\includegraphics[width=4cm]{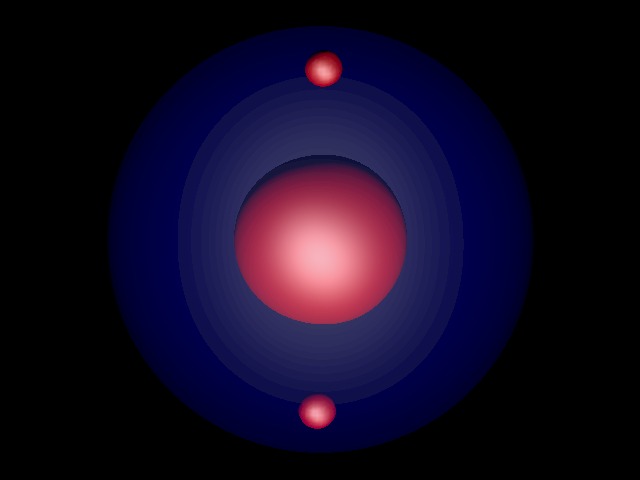}
\includegraphics[width=4cm]{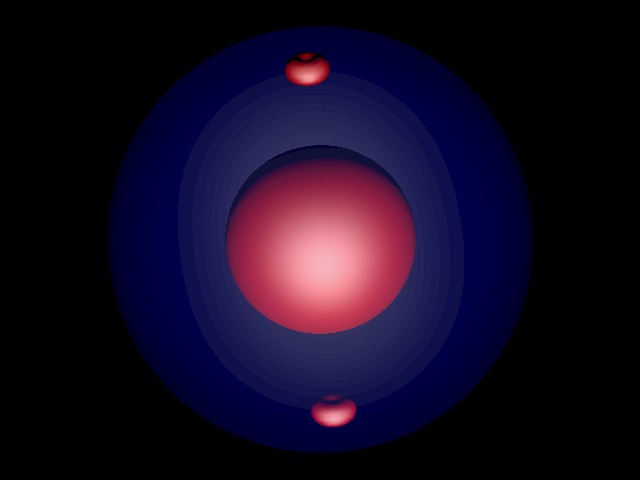}
\includegraphics[width=4cm]{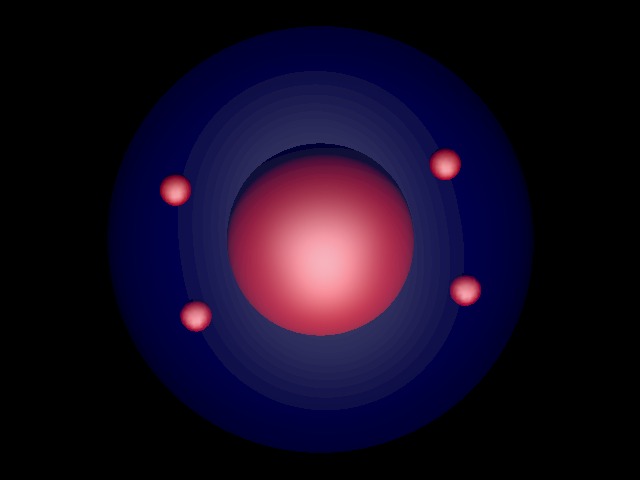}
\includegraphics[width=4cm]{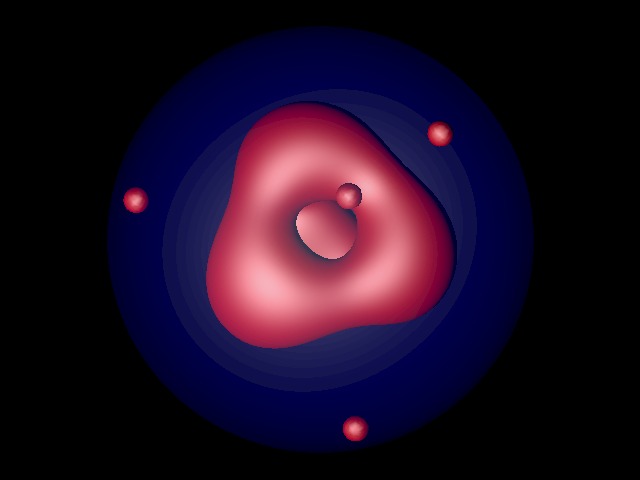}
 \\
\includegraphics[width=4cm]{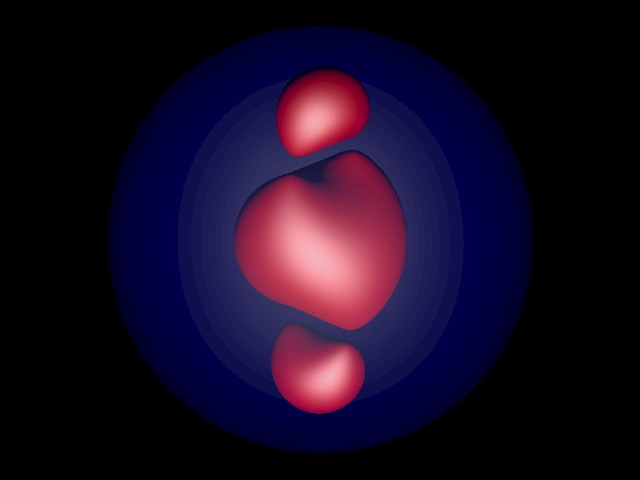}
\includegraphics[width=4cm]{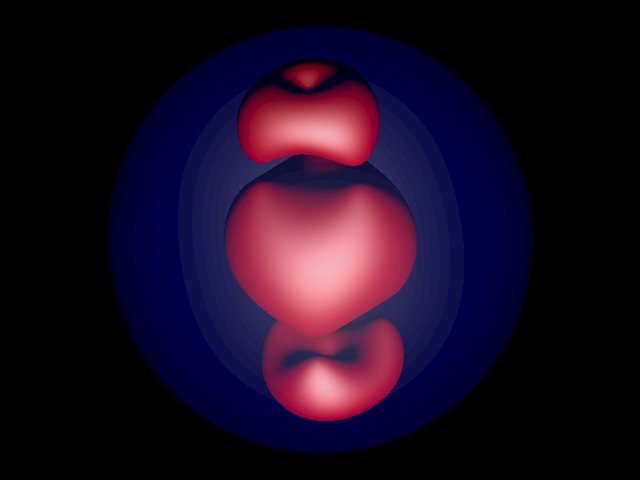}
\includegraphics[width=4cm]{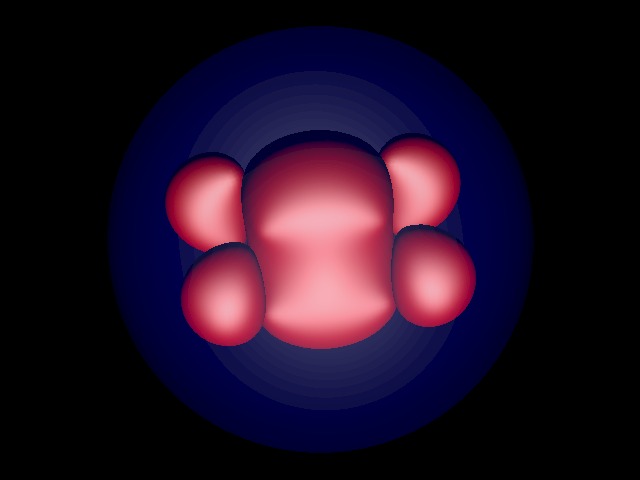}
\includegraphics[width=4cm]{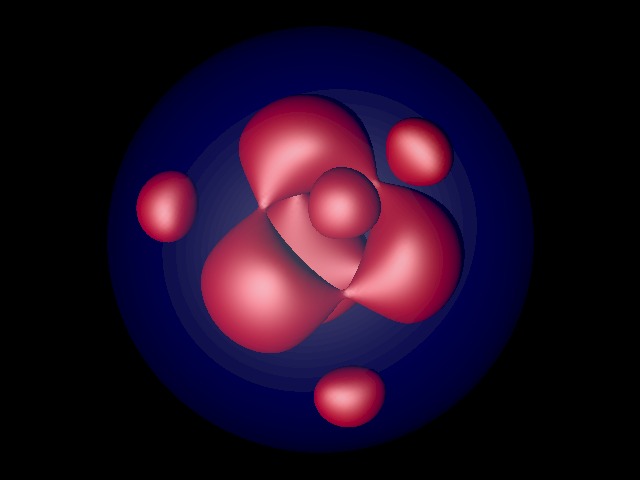}
 \\
\includegraphics[width=4cm]{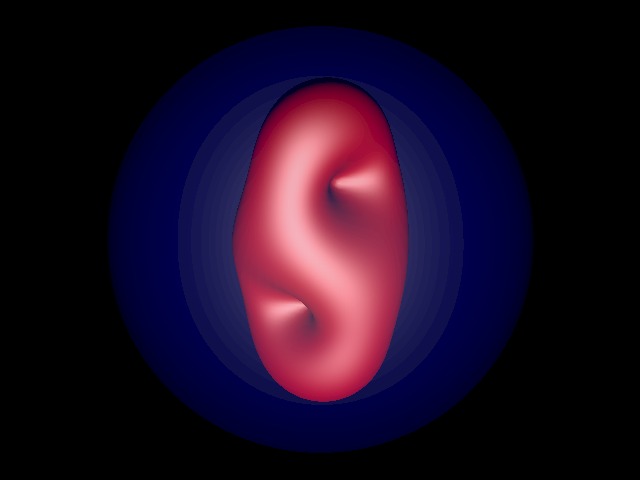} 
\includegraphics[width=4cm]{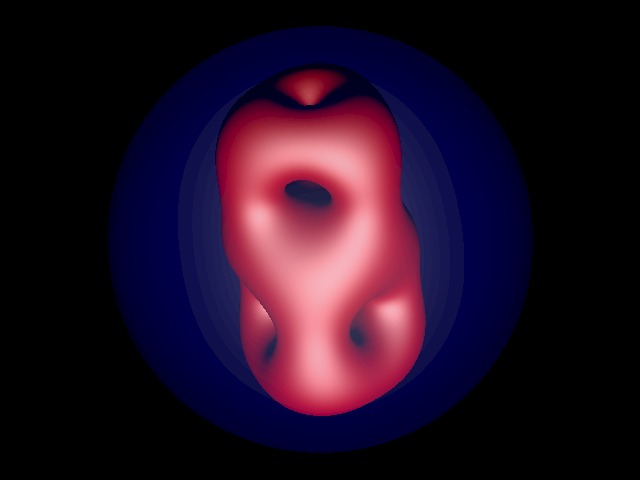}
\includegraphics[width=4cm]{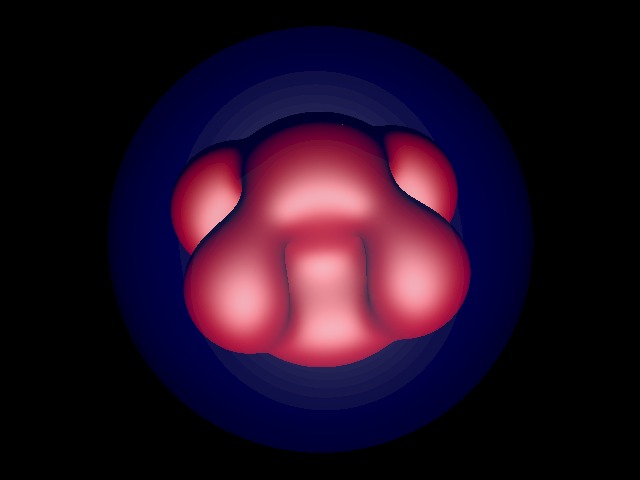}
\includegraphics[width=4cm]{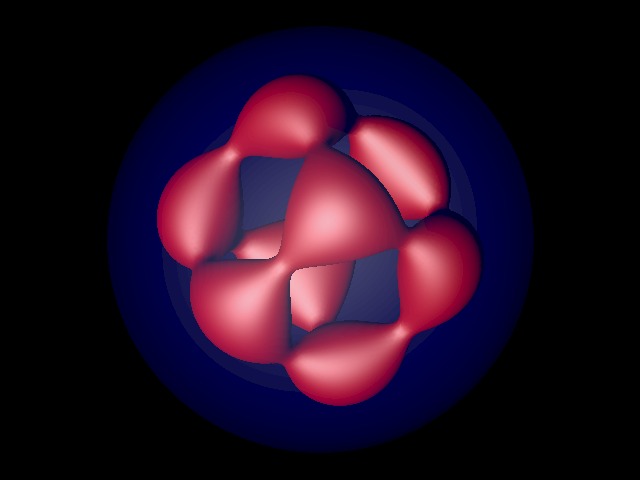}
\\
\includegraphics[width=4cm]{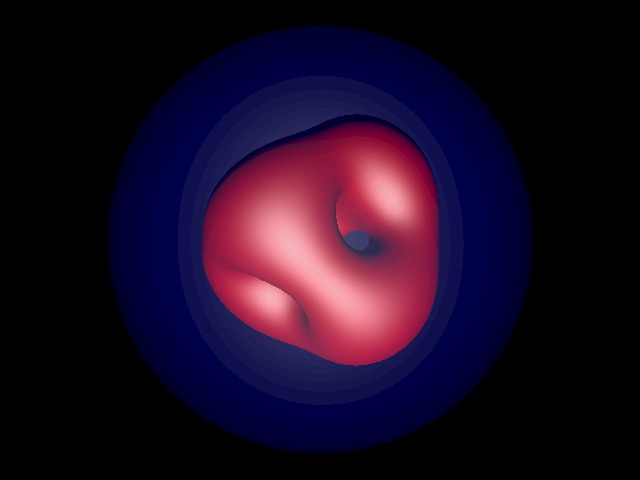} 
\includegraphics[width=4cm]{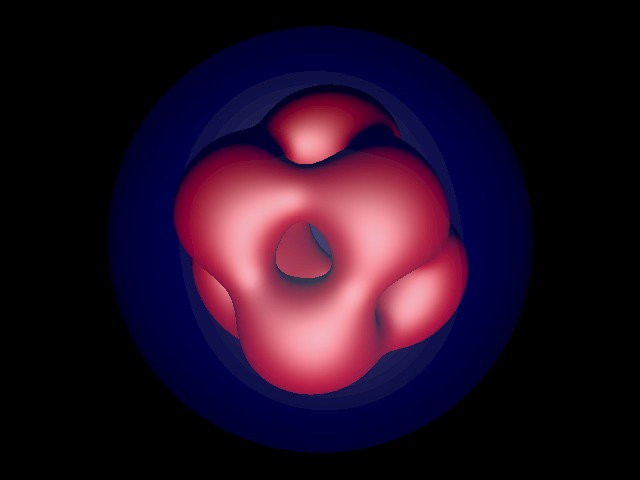}
\includegraphics[width=4cm]{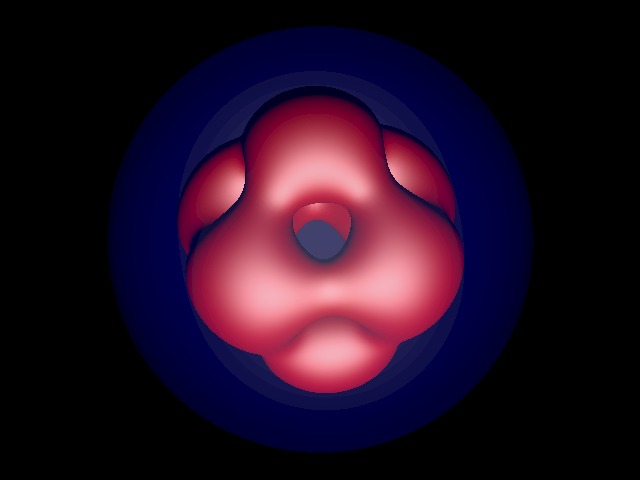}
\includegraphics[width=4cm]{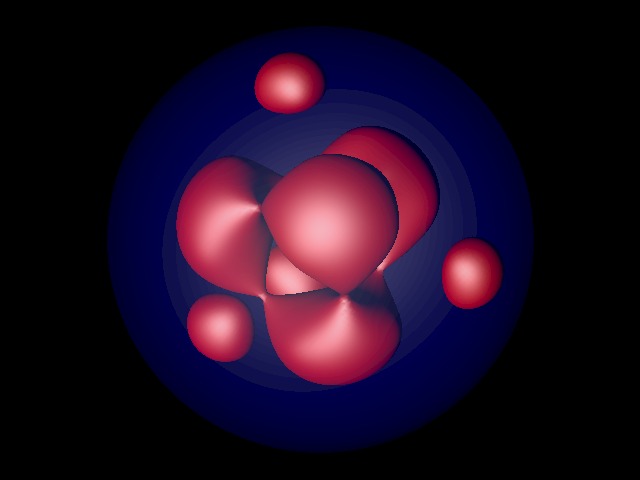}
\\
\includegraphics[width=4cm]{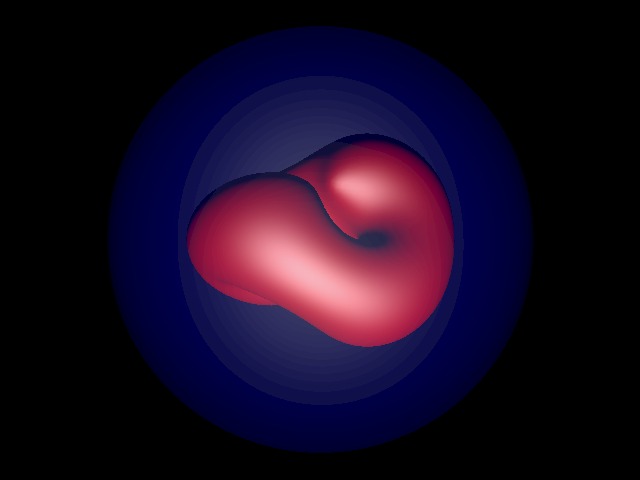} 
\includegraphics[width=4cm]{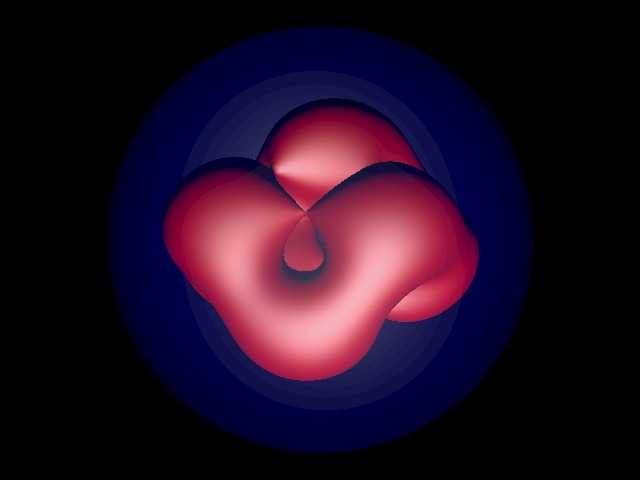}
\includegraphics[width=4cm]{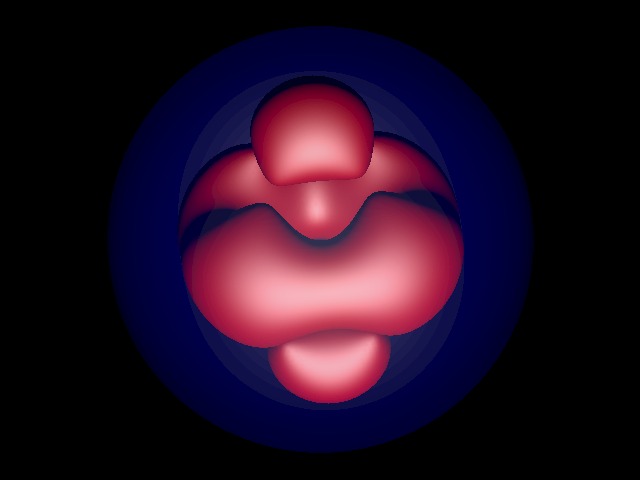}
\includegraphics[width=4cm]{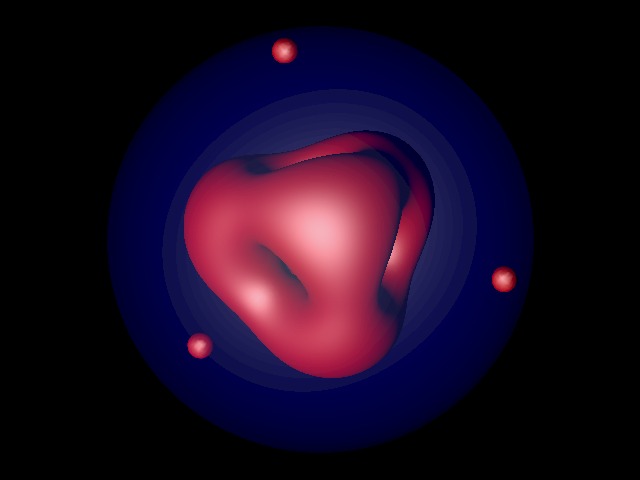}
\\
\includegraphics[width=4cm]{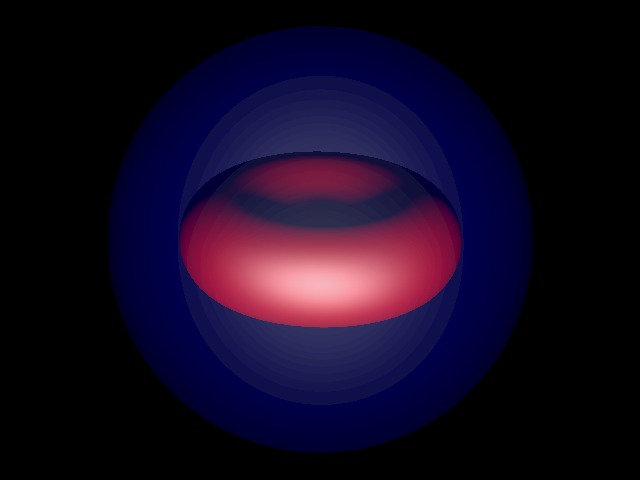} 
\includegraphics[width=4cm]{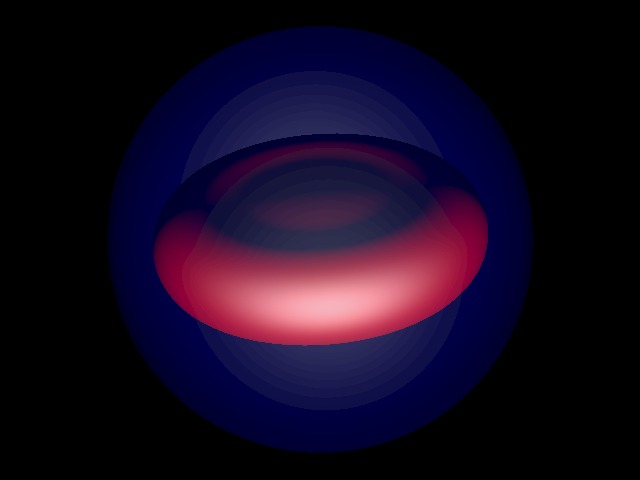}
\includegraphics[width=4cm]{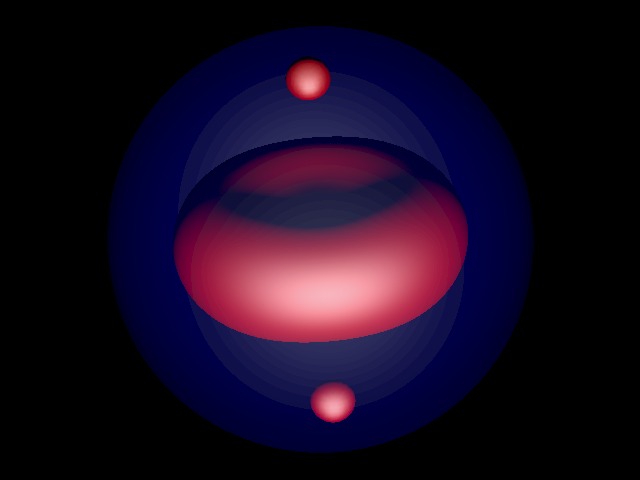}
\includegraphics[width=4cm]{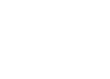}
\\
\caption{Energy density isosurfaces: 
first column $D_2$ symmetric 3-monopoles,
second column $D_3$ symmetric 5-monopoles,
third column $D_4$ symmetric 5-monopoles,
fourth column tetrahedrally symmetric 7-monopoles.
}\label{fig-iso1} 
\end{center}
\end{figure}

\subsection{3-monopoles with $D_2$ symmetry}\label{subsec-3d2}
This example is of the first type, which is perhaps the most obvious method
to construct a symmetric family, since the symmetry of the monopole
is simply the symmetry of the points on the sphere corresponding to 
the positions of the poles.
The one-parameter family is given by $a\in(-1,1)$ where we take the four poles
\be
\xi_0=\sqrt{\frac{1+a}{1-a}}e^{i\pi/4},\
\xi_1=-\xi_0, \ \xi_2=1/\xi_0, \ \xi_3=-1/\xi_0,
\ee
with canonical weights, giving an obvious dihedral $D_2$ symmetry.
The spectral curve is
\be
(\eta-\zeta)(\eta^2+\zeta^2-\frac{4a^2}{1-a^2}\eta\zeta)
+ia(\eta+\zeta)(\eta\zeta+1)(\eta\zeta-1)=0,
\label{d2}
\ee
and is invariant under 
\be
(\eta,\zeta)\mapsto(-\eta,-\zeta),\qquad \qquad
(\eta,\zeta)\mapsto\bigg(\frac{1}{\eta},\frac{1}{\zeta}\bigg),
\label{d2subgroup}
\ee
which generate the $D_2$ subgroup of the tetrahedral group (\ref{tetrotations}).
Note that the change of sign $a\mapsto -a$ 
is equivalent to the $90^\circ$ rotation $(\eta,\zeta)\mapsto(i\eta,i\zeta)$.

The behaviour of the three monopoles as the parameter $a$ is varied can be
determined by an examination of the spectral curve (\ref{d2}) for
particular pertinent values of $a.$
If $a=0$ then (\ref{d2}) becomes the axial curve (\ref{axial}) with $N=3,$
and if $a=1/\sqrt{3}$ it is the tetrahedral curve (\ref{tet3}). 
Given the above comment regarding $a\mapsto -a$ we see that
the curve is also tetrahedrally symmetric if $a=-1/\sqrt{3},$ when the dual
tetrahedron is obtained.
In the limit as $a\to\pm 1$ the spectral curve tends to the curve
$(\eta-\zeta)\eta\zeta=0,$ and we see by comparison with (\ref{star}), that
this is the product of three stars for monopoles with positions
$(X_1,X_2,X_3)$ given by $(0,0,0)$ and $(0,0,\pm 1).$ 
We therefore find that as $a$ is varied from $-1$ to $1$, 
two monopoles from infinity approach a monopole at the origin from opposite
directions along a line, form the tetrahedral 3-monopole, then the
axial 3-monopole, and then separate in the same manner along the same line
but with a $90^\circ$ rotation about the line. 

Using equation (\ref{higgs}) we have an explicit (though cumbersome) expression
for $|\Phi|^2$ for the whole family and hence can obtain an explicit 
form for the energy density by applying the Laplace-Beltrami operator.
This process generates the energy density isosurfaces displayed in
the first column of Figure~\ref{fig-iso1}, which correspond to increasing values
of $a\in(-1,0].$ Plots for $a>0$ are not shown since they are simply 
$90^\circ$ rotations of the plots for $a<0.$
The blue sphere in the energy density plots represents the boundary of 
hyperbolic space and, of course, a monopole appears smaller as it 
approaches this boundary because of the effect of the 
metric in the unit ball model of hyperbolic space.
This one-parameter family is the hyperbolic analogue of the 
twisted line scattering of three Euclidean monopoles presented in \cite{HS3},
where the spectral curve is known via a solution of the Nahm equation but 
the Higgs field and energy density is only available by means of a 
numerical computation of the Nahm transform.

The rational map for this one-parameter family is obtained using equation
(\ref{JNRratmap}) and is given by
\be
{\cal R}=\frac{ia(3+a^2)z^2+1-a^2}{(1-a^2)(z^3+iaz)},
\ee
with the manifest $C_2$ symmetry ${\cal R}(-z)=-{\cal R}(z).$

\subsection{5-monopoles with $D_3$ symmetry}
The 3-monopole twisted line family of subsection \ref{subsec-3d2} includes the 
tetrahedral 3-monopole and there is a similar 5-monopole twisted line 
family that includes the octahedral 5-monopole.
The Euclidean version of this family was identified in \cite{HS3} but 
the associated spectral curves or (numerical) energy density plots have not
been investigated. The hyperbolic version can easily be studied in explicit
detail using our new approach, with the following results.
 
The six canonical weight poles are taken to be
\be
\xi_j=\sqrt{\frac{1+a}{1-a}}e^{i\pi(1+4j)/6}, \quad \xi_{j+3}=1/\xi_j, \quad
j=0,1,2,
\ee
where $a\in(-1,1).$ This yields the spectral curve 
\be
\eta^5-\zeta^5
-\frac{2ia}{\sqrt{1-a^2}}(\eta^5\zeta^3-\eta^3\zeta^5+\eta^2-\zeta^2)
-\frac{(1+3a^2)}{(1-a^2)}(\eta^4\zeta-\eta\zeta^4)
+\frac{(1+10a^2+5a^4)}{(1-a^2)^2}(\eta^3\zeta^2-\eta^2\zeta^3)
=0,
\label{5d3curve}
\ee
which is invariant under the $D_3$ symmetry generated by
\be
(\eta,\zeta)\mapsto(\omega\eta,\omega\zeta),\qquad \qquad
(\eta,\zeta)\mapsto\bigg(\frac{1}{\eta},\frac{1}{\zeta}\bigg),
\label{d3subgroup}
\ee
where $\omega=e^{2\pi i/3}.$
There is octahedral symmetry when $a=-\frac{1}{\sqrt{3}}$ and 
the curve becomes
\be
(\eta-\zeta)\big(\eta^4+\zeta^4-2(\eta^3\zeta+\eta\zeta^3)
+9\eta^2\zeta^2
+\sqrt{2}i(\eta^4\zeta^3+\eta^3\zeta^4+\eta+\zeta)\big)=0,
\ee
which agrees with the earlier octahedral curve (\ref{oct5})
after a spatial rotation.
The replacement $a\mapsto -a$ is equivalent to a rotation through
$60^\circ$ around the main symmetry axis and the
curve (\ref{5d3curve}) is axially symmetric if $a=0.$ 
In the limit $a\to \pm 1$ the curve becomes $(\eta-\zeta)\eta^2\zeta^2=0,$
which is the product of a star for a 1-monopole at the origin and the
curves (\ref{axial2shifted}) for two axial 2-monopoles at infinity with 
positions $(X_1,X_2,X_3)=(0,0,\pm 1).$ This twisted line family therefore
describes two axial 2-monopoles that approach a single monopole at the origin,
from either side of the symmetry axis, form the octahedral 5-monopole, then the axial 5-monopole, with the process then reversing with an 
accompanying rotation by $60^\circ.$ 
Some selected energy density isosurfaces are presented in the 
second column of Figure~\ref{fig-iso1}, for increasing values of $a\in(-1,0].$

The rational map for this family is
\be
{\cal R}=\frac{\sqrt{1-a^2}-4ia\frac{(1+a^2)}{(1-a^2)}z^3}
{\sqrt{1-a^2}z^5-2iaz^2},
\ee
with the $C_3$ symmetry realized as ${\cal R}(\omega z)={\cal R}(z)/\omega^2.$


\subsection{5-monopoles with $D_4$ symmetry}
Our second type of family is perhaps less intuitive then the first type, as
it involves fixing the positions of the poles but varying the weights away
from their canonical values. As an example we present a one-parameter
family of 5-monopoles with $D_4$ symmetry that includes the octahedral 
5-monopole.

The six poles are placed at the vertices of an octahedron
\be
\xi_0=\infty, \ \xi_1=1, \ \xi_2=-1, \ \xi_3=i, \ \xi_4=-i, \ \xi_5=0,
\ee
so this data is of 't Hooft form as one of the poles is at $\infty.$
The weights of the remaining five poles are taken to be
\be \lambda_5^2=1, \qquad
\lambda_1^2=\lambda_2^2=\lambda_3^2=\lambda_4^2,
\ee
 with $\lambda_1\in(0,\infty)$ the
parameter of this family.
If $\lambda_1=\sqrt{2},$ then all weights are canonical and there is
octahedral symmetry, but otherwise the symmetry is broken to $D_4$ symmetry. 

The spectral curve is 
\be
 (\eta-\zeta)
(\eta^4\zeta^4-\eta^4+4\lambda_1^2\eta^3\zeta+4\lambda_1^2\eta\zeta^3-\zeta^4+1)=0,
\label{5d4curve}
\ee
and is invariant under 
\be
(\eta,\zeta)\mapsto(i\eta,i\zeta),\qquad \qquad
(\eta,\zeta)\mapsto\bigg(\frac{1}{\eta},\frac{1}{\zeta}\bigg),
\label{d4subgroup}
\ee
which generate the $D_4$ symmetry.

If $\lambda_1=\sqrt{2}$ then the curve (\ref{5d4curve}) reverts to
the spectral curve (\ref{oct5}) of the octahedral 5-monopole. 
In the limit $\lambda_1\to 0$ the curve (\ref{5d4curve}) becomes
\be
(\eta-\zeta)(\eta^4\zeta^4-\eta^4-\zeta^4+1)=0=
(\eta-\zeta)\prod_{j=1}^4(\eta+i^j)(\zeta-i^j)
\ee
which is the product of stars for five monopoles, with one
at the origin $(X_1,X_2,X_3)=(0,0,0)$ 
and the remaining four monopoles at the boundary of hyperbolic space
along the Cartesian axes $(\pm 1,0,0)$ and $(0,\pm 1,0).$
In the opposite limit $\lambda_1\to\infty$ the curve becomes
\be
\eta\zeta(\eta^3-\eta^2\zeta+\eta\zeta^2-\zeta^3)=0,
\ee
where the first two factors describe 1-monopoles at the boundary
of hyperbolic space with positions 
$(0,0,\pm 1)$ and the final factor is the 
spectral curve of the axial 3-monopole at the origin.

We therefore see that as $\lambda_1$ increases through the
 interval $(0,\infty)$, four 1-monopoles approach from infinity along
the Cartesian axes in the plane $X_3=0$ and merge with a 1-monopole
at the origin to form the octahedral 5-monopole. The octahedral 5-monopole
then splits to produce two 1-monopoles moving in opposite directions along
the $X_3$-axis, leaving behind the axial 3-monopole.
Corresponding energy density isosurfaces are displayed in the third column
of Figure~\ref{fig-iso1}. Note that, as with some of the other energy
density isosurfaces presented in this paper, we often slightly 
rotate the image to obtain an improved viewing angle, so for example the 
$X_3$-axis may not be exactly aligned with the vertical, although the images
within each column have the same viewing angle.

 The rational map for this one-parameter family is 
\be
{\cal R}=\frac{(4\lambda_1^2+1)z^4-1}{z^5-z},
\ee
with the clear $C_4$ symmetry ${\cal R}(iz)=-i{\cal R}(z).$

\subsection{7-monopoles with tetrahedral symmetry}
Our next example of a family of the second type illustrates the fact
that dihedral symmetry, although convenient for producing one-parameter
families, is certainly not the only possibility. In this subsection, 
we construct a one-parameter family
of 7-monopoles by imposing tetrahedral symmetry. 

The eight poles are taken to be the 
roots of the Klein polynomial (\ref{kleinoctf}) for the face centres of the
octahedron (or equivalently the vertices of the cube).
Explicitly, we label the poles as
\be
\xi_0=\frac{1+i}{\sqrt{3}+1}, \ \xi_1=-\xi_0, \ \xi_2=\xi_0^{-1}, \ \xi_3=-\xi_0^{-1}, \
\xi_4=\frac{1-i}{\sqrt{3}+1}, \ \xi_5=-\xi_4, \ \xi_6=\xi_4^{-1}, \ \xi_7=-\xi_4^{-1},
\ee
and take $\lambda_j^2$ to be canonical weights for $j=0,1,2,3$ but 
$\mu^2$ times the canonical weights for $j=4,5,6,7.$
The one-parameter family is given by $\mu\in(0,\infty)$ with
the resulting spectral curve taking the form
\bea 
& &
\bigg((\eta-\zeta)^3+\frac{i}{\sqrt{3}}(\eta+\zeta)(\eta\zeta+1)(\eta\zeta-1)
\bigg)
\prod_{j=4}^7(\eta\zeta\bar\xi_j+\zeta-\eta|\xi_j|^2-\xi_j)
\\
& &+\mu^2
\bigg((\eta-\zeta)^3-\frac{i}{\sqrt{3}}(\eta+\zeta)(\eta\zeta+1)(\eta\zeta-1)
\bigg)
 \prod_{j=0}^3(\eta\zeta\bar\xi_j+\zeta-\eta|\xi_j|^2-\xi_j)=0,
\nonumber
\eea
where the first term is the product of the tetrahedral 3-monopole curve
(\ref{tet3}) and four stars for monopoles on the sphere at infinity 
on the vertices of the dual tetrahedron. The second term is $\mu^2$ times
the first term after the replacement $(\eta,\zeta)\mapsto(i\eta,i\zeta).$
The transformation $\mu\mapsto \mu^{-1}$ is therefore equivalent to a rotation
by $90^\circ$ around the $X_3$-axis. 

If $\mu=1$ then the tetrahedral symmetry is enhanced to cubic symmetry, as
there are eight poles with canonical weights on the vertices of a cube. 
$N=7$ is not the lowest value of $N$ for which there is a hyperbolic monopole
with cubic symmetry. The lowest value is $N=4$ and the spectral curve can
be found in \cite{NoRo} with the explicit Higgs field derived in \cite{MS}
using the ADHM construction with circle invariance. However, this cubic
4-monopole is clearly not within the JNR class, as five points cannot be
placed on a sphere with cubic symmetry.  

We see from the above spectral curve 
that as $\mu$ increases through the interval $(0,\infty),$ four monopoles
approach from infinity towards the face centers of the tetrahedral 3-monopole.
The monopoles then merge to form a cubic 7-monopole which subsequently
splits to leave the dual tetrahedral 3-monopole with four monopoles receding 
from the face centres towards infinity. Energy density isosurfaces are 
displayed in the fourth column of Figure~\ref{fig-iso1} for increasing
values of $\mu.$ 

For values of $\mu$ around that associated with the
second image in the fourth column of Figure~\ref{fig-iso1} 
(or equivalently the fourth image in this column), we may view this
solution as a prototype hyperbolic analogue of the multi-shell 
Euclidean monopoles suggested in \cite{Man} within the magnetic bag 
approximation.

The rational map for the family is
\be
{\cal R}=\frac{
(1-\mu^4)(5z^6-z^2)
-i\sqrt{3}(1+\mu^2)^2(11z^4+1)}
{i\sqrt{3}(1+\mu^2)^2(z^7+3z^3)
+(1-\mu^4)(5z^5+z)},
\ee
with the evident $C_{2}$ symmetry ${\cal R}(-z)=-{\cal R}(z).$
For the cubic $\mu=1$ case the map simplifies to 
\be
{\cal R}=\frac{11z^4+1}{z^7+3z^3},
\ee
with the manifest $C_4$ symmetry ${\cal R}(iz)=i{\cal R}(z).$

\begin{figure}
\begin{center}
\includegraphics[width=4cm]{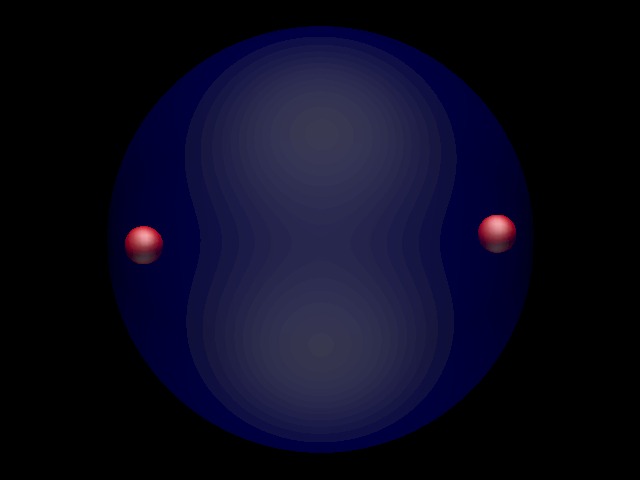}
\includegraphics[width=4cm]{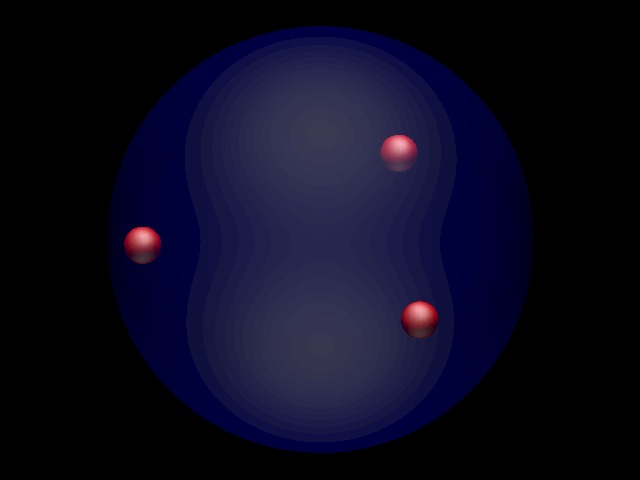}
\includegraphics[width=4cm]{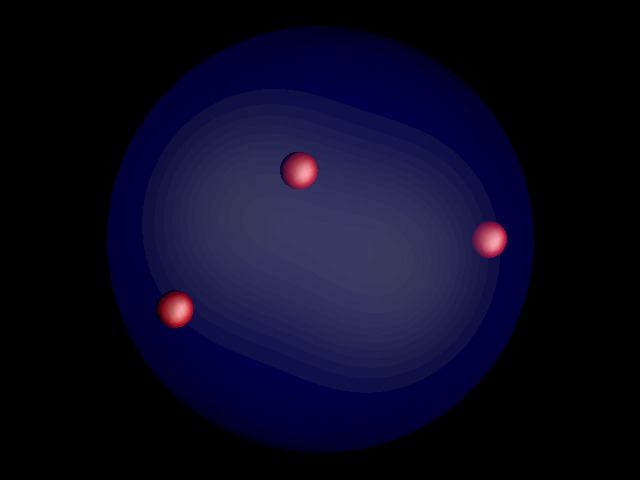}
 \\
\includegraphics[width=4cm]{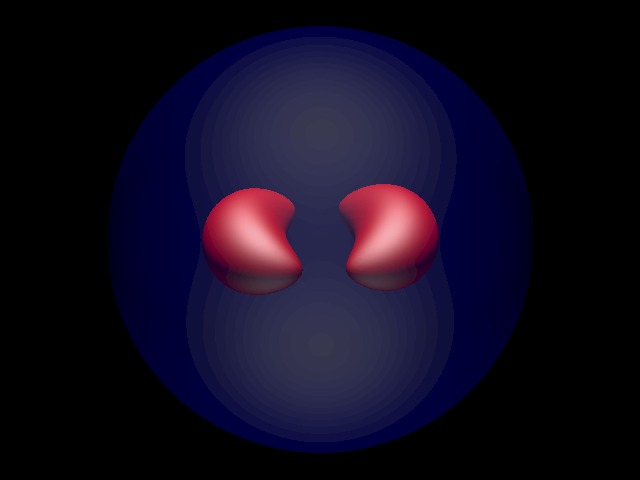}
\includegraphics[width=4cm]{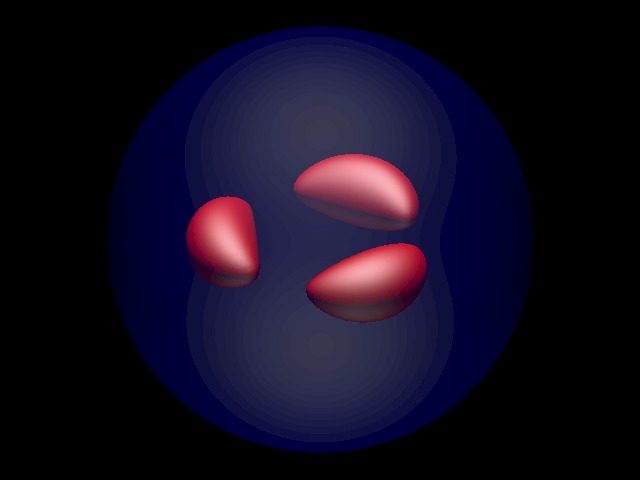}
\includegraphics[width=4cm]{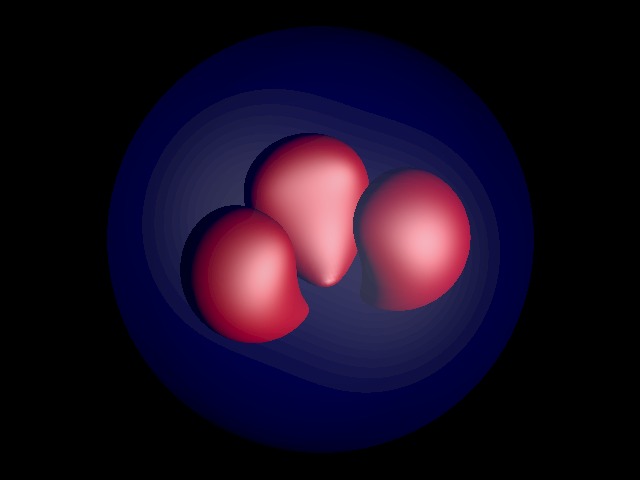}
 \\
\includegraphics[width=4cm]{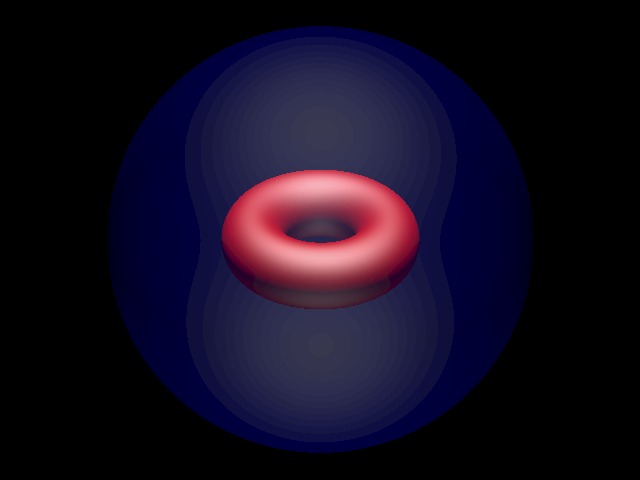}
\includegraphics[width=4cm]{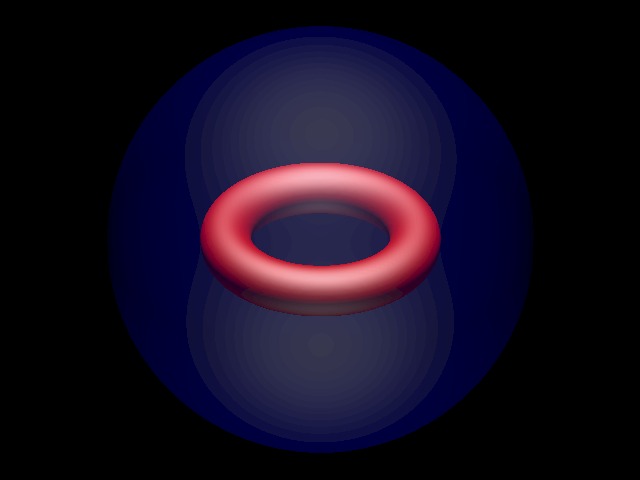}
\includegraphics[width=4cm]{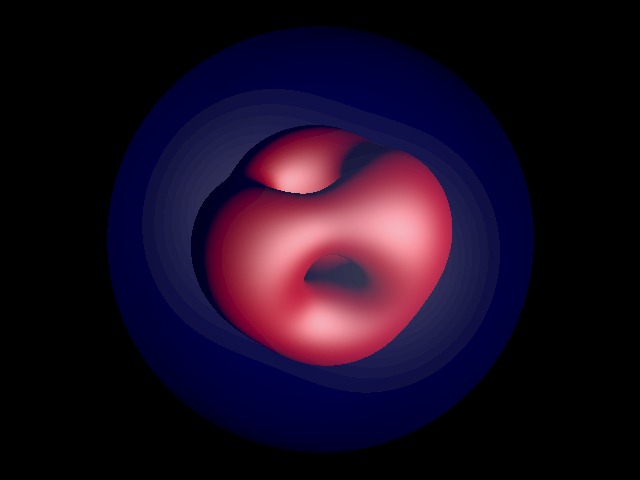}
 \\
\includegraphics[width=4cm]{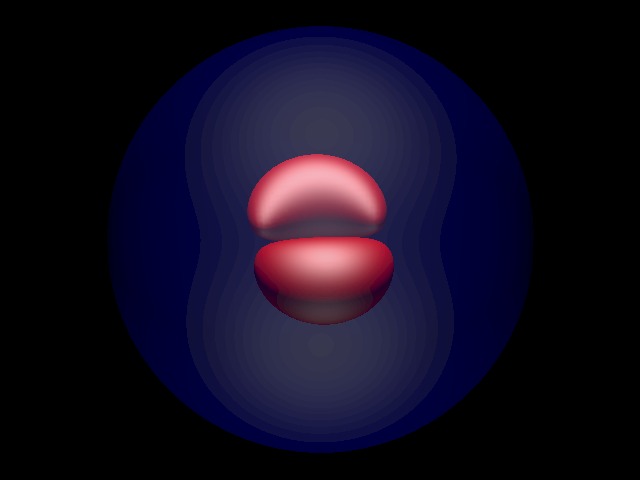}
\includegraphics[width=4cm]{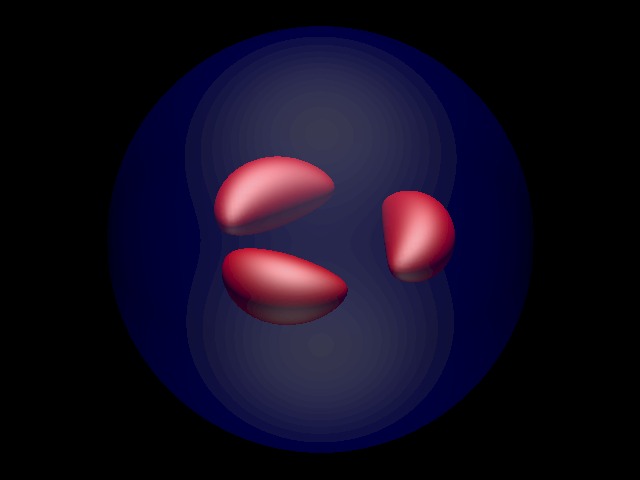}
\includegraphics[width=4cm]{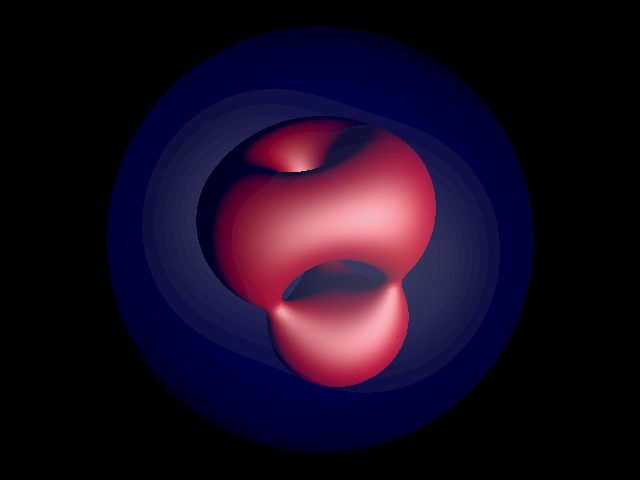}
 \\
\includegraphics[width=4cm]{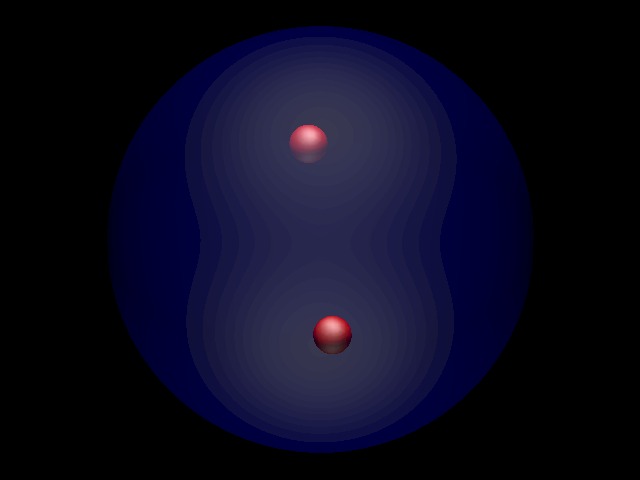}
\includegraphics[width=4cm]{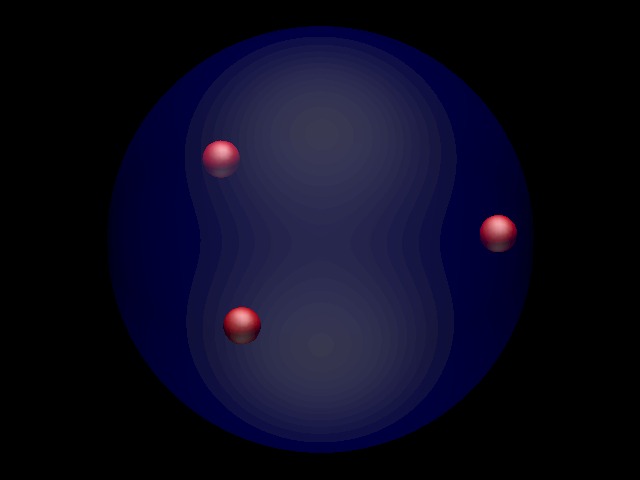}
\includegraphics[width=4cm]{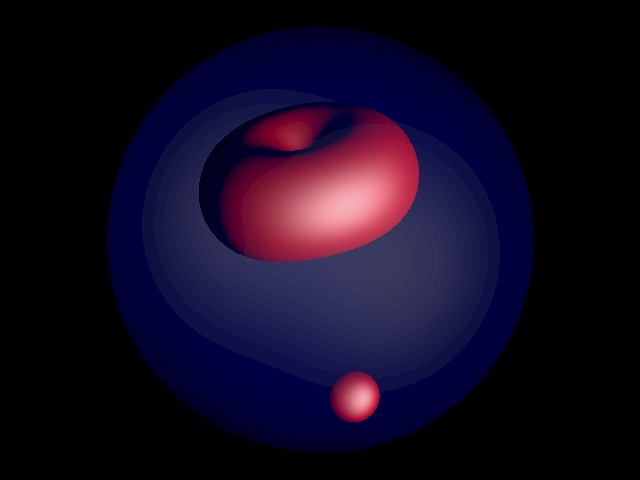}
 \\
\caption{Energy density isosurfaces: 
first column $D_2$ symmetric 2-monopoles,
second column $D_3$ symmetric 3-monopoles,
third column $C_3$ symmetric 3-monopoles.
}\label{fig-iso2} 
\end{center}
\end{figure}

\subsection{2-monopoles with $D_2$ symmetry}
Our first example of a family of type three, where the positions of the
poles vary together with the (generically) non-canonical weights, is a 
one-parameter family of $D_2$ symmetric 2-monopoles. Although this is
perhaps the simplest family of multi-monopoles, and was studied in 
\cite{MS} using the ADHM formalism, 
its analysis in terms of the JNR approach is a little subtle, and
is therefore worth presenting. 

It is not immediately obvious how to place three poles 
and select their weights so that there is $D_2$ symmetry.
Clearly this requires exploiting the degeneracy that arises when 
all poles lie on a circle, which we take to be the unit circle
in the plane $X_3=0.$ 
Imposing the subgroup $C_2$ symmetry given by a rotation by
$180^\circ$ around the $X_1$-axis is straightforward, as one of the
poles can be placed on the $X_1$-axis with the two remaining poles
placed symmetrically about the axis with equal weights. 
Explicitly, let $a\in(-1,1)$ be the parameter of the family and
set
\be
\xi_0=1, \ \ \xi_1=\frac{a-1}{2}+\frac{i}{2}\sqrt{3+2a-a^2}, 
\ \ \xi_2=\xi_1^{-1},  
\ \ \lambda_1^2=\lambda_2^2=1,
\ee
with the weight $\lambda_0^2$ undetermined for the moment.

The spectral curve is
\be
\frac{(\lambda_0^2+a-1)}{(2+\lambda_0^2)}
(\eta^2\zeta^2+1)
+\frac{(\lambda_0^2-a\lambda_0^2-a-1)}{(2+\lambda_0^2)}
(\eta^2\zeta-\eta\zeta^2-\eta+\zeta)
+\eta^2+\zeta^2
-\frac{(2+2a+\lambda_0^2(1-a)^2)}{(2+\lambda_0^2)}
\eta\zeta=0,
\label{tmp2d2curve}
\ee
and is invariant under the  $C_2$ symmetry 
$(\eta,\zeta)\mapsto({\eta}^{-1},{\zeta}^{-1}).$
This $C_2$ symmetry is extended to $D_2$ symmetry by 
requiring invariance of the spectral curve
(\ref{tmp2d2curve}) under the 
additional generator $(\eta,\zeta)\mapsto(-\eta,-\zeta).$ 
This extra symmetry requires that the
coefficient $c_{ij}$ vanishes unless $(i+j)\, \mbox{mod}\, 2=0.$
This is satisfied providing 
\be
\lambda_0^2=\frac{1+a}{1-a},
\label{mun2}
\ee
which yields the required $D_2$ invariant spectral curve
\be
\eta^2+\zeta^2 +a(\eta^2\zeta^2+1)
+(a^2-1)\eta\zeta=0.
\label{2d2}
\ee
We see from (\ref{2d2}) that $a\mapsto -a$ is equivalent to the 
$90^\circ$ rotation $(\eta,\zeta)\mapsto(i\eta,i\zeta),$ and 
furthermore the
axial 2-monopole curve is recovered by setting $a=0.$

The boundary of hyperbolic space intersects the plane $X_3=0$ in the
circle given by 
$(X_1,X_2,X_3)=(\cos\theta,\sin\theta,0)$, and from (\ref{star})
 a monopole at this position corresponds to the star
\be
(\eta+e^{i\theta})(\zeta-e^{i\theta})=0.
\ee
In the limit $a\to-1$ the curve (\ref{2d2}) becomes the product of two stars
\be
(\eta+1)(\zeta-1)(\eta-1)(\zeta+1)=0
\label{stars1inf}
\ee
for monopoles with positions $(\pm 1,0,0).$ Therefore as $a$ increases through
$(-1,1)$ the two monopoles approach along the $X_1$-axis, merge to form the
axially symmetric 2-monopole, and separate along the $X_2$-axis. This is the
hyperbolic analogue of the famous right angle scattering of two 
Euclidean monopoles discovered by Atiyah and Hitchin \cite{AH}. 
Energy density isosurfaces are displayed for increasing values of $a$ in the
first column of Figure~\ref{fig-iso2}.

The rational map for this family is 
\be
{\cal R}=\frac{1-a^2}{z^2+a},
\label{map2d2}
\ee
with the manifest $C_2$ symmetry ${\cal R}(-z)={\cal R}(z).$

\subsection{3-monopoles with $D_3$ symmetry}
The one-parameter family of $D_2$ symmetric 2-monopoles, 
studied in the previous subsection, has a
generalization to a one-parameter family of $D_N$ symmetric $N$-monopoles.
The $N$ monopoles are located on the vertices of a contracting
regular $N$-gon, merge to form the axial $N$-monopole, and then separate
on the vertices of an expanding regular $N$-gon, that is obtained from the
incoming polygon by a rotation through $180^\circ/N.$

We illustrate this generalization by presenting the result 
for $N=3$.
The four poles are again taken to lie on the unit circle, but this time
two of the poles are placed on the $X_1$-axis to achieve the 
$C_2$ symmetry given by a rotation by $180^\circ$ around the $X_1$-axis.
As before, the two remaining poles are
placed symmetrically about this axis with equal weights.
In detail, the parameter is $a\in(-1,1)$ and the poles and weights are 
\be
\xi_0=1,\ 
\xi_1=-1,\ 
\xi_2=\frac{1}{2}(a+i\sqrt{4-a^2}),\
\xi_3=\xi_2^{-1}, \
\lambda_0^2=1, \ \lambda_2^2=\lambda_3^2,
\ee
with $\lambda_1$ and $\lambda_2$ yet to be determined.
The $D_3$ symmetry is obtained by demanding that the spectral curve is
invariant under the additional $C_3$ symmetry 
$(\eta,\zeta)\mapsto(\omega \eta,\omega \zeta),$ where $\omega=e^{2i\pi/3}.$
This results in the requirement that
$c_{ij}=0$ if $(i+j)\, \mbox{mod}\, 3\ne 0,$
which gives
\be
\lambda_1^2=\frac{(1-a)(2-a)}{(1+a)(2+a)}, \qquad
\lambda_2^2=\frac{2(1-a)}{(2+a)}.
\ee
The one-parameter family of $D_3$ symmetric spectral curves is then
\be
\eta^3-\zeta^3+a(\eta^3\zeta^3-1)+(a^2-1)(\eta^2\zeta-\eta\zeta^2)=0,
\ee
which satisfies all the properties expected of this family, as described
at the start of this subsection. Some energy density isosurfaces are
displayed in the second column of Figure~\ref{fig-iso2} for increasing
values of $a\in(-1,1).$  

The rational map is
\be
{\cal R}=\frac{1-a^2}{z^3+a},
\label{rat3d3}
\ee
being the obvious generalization of (\ref{map2d2}). For larger values of
$N$ the procedure follows the same process as in this subsection and the 
previous one, with one pole on the $X_1$-axis if $N$ is even and two if
$N$ is odd. The remaining poles are placed symmetrically in pairs 
around the unit circle, with
equal weights to attain the $C_2$ symmetry, with the weights 
then determined
by applying an additional $C_N$ generator to enforce the full $D_N$
symmetry. 
\subsection{3-monopoles with $C_3$ symmetry}
Our final example illustrates a phenomenon that appears if cyclic
symmetry is imposed, rather than dihedral symmetry. Imposing cyclic 
symmetry will produce more than a one-parameter family, as there will be
an additional degree of freedom associated with a translation of the
whole configuration along the symmetry axis. 

In Euclidean space,
the motion of $N$ monopoles has a natural decomposition into a 
trivial motion of the centre of mass of the configuration and a 
non-trivial relative motion between monopoles. In terms of the moduli space
approximation, this allows (without loss of generality) a 
restriction to centered monopoles, 
in which the centre of mass is fixed at the origin.
In hyperbolic space the situation is not so simple, since there is no 
definition of the centre of mass (even for point particles) that has all
the properties that exist in the Euclidean setting.
Fortunately there is a definition \cite{MNS} for a hyperbolic monopole to
be centered, so we will apply this definition to restrict to a one-parameter
family. 

The definition introduced in \cite{MNS} for a centered hyperbolic monopole is
by no means obvious and relies upon the use of yet another holomorphic object
associated with a hyperbolic monopole, namely a holomorphic sphere.
This is a holomorphic map from $\mathbb{CP}^1$ to $\mathbb{CP}^N$, and the
action of the isometries of hyperbolic space induces a moment map whose
zero set can be used to define conditions for a spectral curve to 
correspond to a centered hyperbolic monopole. These conditions map to
simple linear relations between the coefficients $c_{ij}$ in the spectral
curve. All the spectral curves that we have presented so far obey
these centered conditions, as a result of the symmetries that we have imposed.  
In particular, rewriting the results of \cite{MNS} in terms of
these linear relations we find that a 3-monopole is centered if the
coefficients of its spectral curve satisfy 
\be
9c_{30}-c_{21}-c_{12}+9c_{03}=3c_{31}-2c_{22}+3c_{13}=0.
\label{cen3}
\ee
We shall make use of this condition shortly.

The cyclic example we consider is $C_3$ symmetric 3-monopoles obtained
from the following choice of four poles,
\be
\xi_0=0, \qquad  \xi_j=\sqrt{\frac{1+a}{1-a}}\omega^j 
\quad \mbox{ for } j=1,2,3,
\ee 
where $\omega=e^{2\pi i/3}$ and $a\in(-1,1)$ is the free parameter.
The weights $\lambda_j^2$ are canonical for $j=1,2,3$ but 
$\lambda_0^2$ is free for the moment.
This yields the two-parameter family of $C_3$ symmetric spectral curves 
\be
\lambda_0^2\sqrt{\frac{1+a}{1-a}}(\eta^3\zeta^3-1)-\lambda_0^2\frac{(1+a)^2}{(1-a)^2}\eta^3
+\frac{6(1+a)}{(1-a)^2}\eta^2\zeta
-\frac{6}{(1-a)}\eta\zeta^2+\frac{(6+\lambda_0^2-a\lambda_0^2)}{(1+a)}\zeta^3
=0,
\label{3c3}
\ee
invariant under the symmetry $(\eta,\zeta)\mapsto(\omega\eta,\omega\zeta).$

We now reduce this two-parameter family to a one-parameter family 
by imposing the centered condition (\ref{cen3}), which determines the
weight to be
\be
\lambda_0^2=\frac{9-20a+7a^2}{3a(3+a^2)}.
\label{weightcen}
\ee
The requirement that $\lambda_0^2>0$ imposes the restriction 
$a\in(0,a_\star)$, where $a_\star=(10-\sqrt{37})/7.$

Substituting (\ref{weightcen}) into (\ref{3c3}) 
produces the centered spectral curve 
\be
\sqrt{\frac{1+a}{1-a}}(\eta^3\zeta^3-1)-\frac{(1+a)^2}{(1-a)^2}\eta^3
+\frac{18a(3+a^2)
(\eta^2\zeta(1+a)-\eta\zeta^2(1-a))
}{(1-a)^2(9-20a+7a^2)}
+\frac{(9+16a+11a^2)}{(9-20a+7a^2)}\zeta^3
=0.
\label{final3c3}
\ee
The curve has tetrahedral symmetry if $a=\frac{1}{3}$ when it becomes
\be\sqrt{2}\eta^3\zeta^3-4\eta^3+18\eta^2\zeta-9\eta\zeta^2+5\zeta^3-\sqrt{2}=0,
\ee
with the extra $C_2$ symmetry
\be
(\eta,\zeta)\mapsto\bigg(\frac{\sqrt{2}-\eta}{\sqrt{2}\eta+1},
\frac{\sqrt{2}-\zeta}{\sqrt{2}\zeta+1}
\bigg).
\ee
This tetrahedral curve is equal to the earlier tetrahedral curve
(\ref{tet3}) after a suitable rotation. 
In the limit $a\to 0$ the curve is a product of stars
\be
\eta^3\zeta^3-\eta^3+\zeta^3-1=0=\prod_{j=1}^3(\eta+\omega^j)(\zeta-\omega^j)
\ee
for three monopoles on the vertices of an equilateral triangle 
in the plane $X_3=0$ at the boundary of hyperbolic space.
In the limit $a\to a_\star$ the curve becomes
\be
\zeta\bigg(
(1+b)^4\eta^2+(1-b)^4\zeta^2-{(1-b^2)^2}\eta\zeta
\bigg)=0,
\ee
where $b$ is given by the relation $a_\star=2b/(1+b^2),$ 
so $b=(7-2\sqrt{5\sqrt{37}-22})/(10-\sqrt{37})\approx 0.3.$
This curve is the product of a star for a monopole  
at $(0,0,-1)$ and the curve (\ref{axial2shifted}) for an axial 2-monopole
at $(0,0,b).$
The interesting new phenomenon here is that the single monopole is at
infinity when the axial 2-monopole is at a finite distance from the
origin, despite the fact that the total configuration is centered.
This contrasts with the Euclidean situation, where an $N$-monopole cannot
be centered if it consists of two clusters with only one cluster at infinity,
as is self-evident from the properties of the 
centre of mass in Euclidean space. 

A possible physical understanding of this new phenomenon in hyperbolic space 
is that the condition for a hyperbolic monopole to be centered should
be similar to a requirement that the magnetic field on the sphere at 
infinity has a vanishing dipole. A definition of this sort would be 
quite natural, given that the abelian magnetic field on the 
sphere at infinity completely determines the monopole \cite{BA}. 
A single monopole has a finite dipole even as its position tends to the sphere
at infinity in hyperbolic space, so this can indeed be cancelled by a non-zero 
dipole of a cluster in the interior of hyperbolic space. At the moment this
is nothing more than an attempt at a potential physical understanding of
this surprising phenomenon, but it at least suggests why the result is
not unreasonable.  
 
In summary, the one-parameter family described in this subsection consists
of three monopoles that approach on the vertices of a contracting triangle,
merge to form the tetrahedral 3-monopole, which then splits into a single
monopole that travels down the symmetry axis of the triangle, leaving an
axial 2-monopole at a finite distance up the symmetry axis. A selection of
the corresponding energy density isosurfaces are displayed in the third
column of Figure~\ref{fig-iso2}. This process is a hyperbolic analogue
of the $C_3$ symmetric scattering of three Euclidean monopoles, for which
similar energy density isosurfaces can be seen in \cite{Su}; except that
the axial 2-monopole continues to travel along the axis. These Euclidean
results involve a numerical computation of the relevant solution
of the Nahm equation, as the associated genus four curve is the Galois cover
of a genus two curve (rather than an elliptic curve). Progress has
been made in computing the spectral curve for this Euclidean case
\cite{BDE}, but it is significantly more complicated than the hyperbolic
spectral curve (\ref{final3c3}).

The rational map for the centered $C_3$ symmetric 3-monopole family is
\be
{\cal R}=\frac{18a(3+a^2)\sqrt{\frac{1+a}{1-a}}z}
{(1-a)(11a^2+16a+9)z^3-\sqrt{1-a^2}(7a^2-20a+9)},
\ee
with the symmetry realized as ${\cal R}(\omega z)=\omega {\cal R}(z)$.
Note that this is a different realization of the $C_3$ symmetry
than for the rational map (\ref{rat3d3}) of the $D_3$ symmetric
3-monopole of the previous subsection, 
where ${\cal R}(\omega z)={\cal R}(z).$ 
Although both families involve three monopoles on the vertices
of a contracting triangle, the subsequent different configurations are
a result of different arrangements of the relative phases, which are
captured by the above rational map realizations of the $C_3$ symmetry. 

\section{A metric on the space of JNR data}\label{sec-metric}\quad\
Monopole dynamics in Euclidean space can be approximated by geodesic
motion on the monopole moduli space equipped with the
natural $L^2$ metric \cite{Ma1}. 
However, the $L^2$ metric on the moduli space of hyperbolic monopoles 
is infinite, so this approximation is unavailable for studying the
dynamics of hyperbolic monopoles. An alternative approach is to use the
fact that a hyperbolic monopole is uniquely determined by its abelian magnetic
field on the boundary of hyperbolic space, and the associated abelian
connection can be used to define a finite metric on the monopole moduli space
\cite{BA}.
This metric is invariant under the isometries of hyperbolic space,
and in the case of a single monopole the moduli space equipped with
this metric is simply hyperbolic space itself. Applying the moduli space
approximation with this metric therefore yields the natural result that
a slowly moving single hyperbolic monopole follows a geodesic in hyperbolic
space.  

In this section we provide a simple integral formula for the above metric
restricted to the space of JNR data and 
illustrate its application by explicit computation to confirm 
that hyperbolic space is obtained as the moduli space for a single monopole.

To present the metric it is most convenient to use the upper half space model
of hyperbolic space, where the boundary is given by $r=0$ and we set
$z=x_1+ix_2$ to be the complex coordinate on the boundary.
As shown in \cite{MNS}, the required connection on the sphere at
infinity can be written in terms of a hermitian metric obtained by evaluating
the spectral curve on the antidiagonal. Explicitly, the abelian connection
${\cal A}_z=\frac{1}{2}({\cal A}_1-i{\cal A}_2)$ 
is given in terms of the hermitian metric $h(z,\bar z)$ by
\be
{\cal A}_z=\frac{1}{2}\partial_z \log h,
\ee
where $h(z,\bar z)$ is the polynomial in $z$ and $\bar z$ obtained 
as $\bar z^N$ times the spectral curve evaluated
on the antidiagonal $\zeta=z$ and $\eta=-1/\bar z.$ 
Using (\ref{JNRspectralcurve}) gives the hermitian metric in terms of the
JNR data as
\be
h(z,\bar z)=
\sum_{j=0}^N\lambda_j^2\mathop{\prod_{k=0}^N}_{k\ne j}|z-\gamma_k|^2
=\psi|_{r=0}
\prod_{k=0}^N|z-\gamma_k|^2.
\label{hmetric}
\ee
Let $t_\mu$ for $\mu=1,\ldots,\mbox{dim}({\mathbb{M}}^{\mbox{\tiny{JNR}}}_N)$
be real independent coordinates on the JNR moduli space.
The metric is the $L^2$ metric of the abelian connection
\be
g_{\mu\nu}=K\int \frac{\partial{\cal A}_i}{\partial t_\mu}\, 
\frac{\partial{\cal A}_i}{\partial t_\nu}\, 
\,d^2x
=K\int
\bigg(\frac{\partial}{\partial t_\mu}\bigg(\frac{\partial_i h}{h}\bigg)\bigg)
\bigg(\frac{\partial}{\partial t_\nu}\bigg(\frac{\partial_i h}{h}\bigg)\bigg)
\,d^2x,
\ee
where $K$ is a normalization constant.

As an example, consider the case $N=1,$ where the three real 
independent coordinates may be taken to be those in the 't Hooft data,
that is,
$t_1+it_2=\gamma_1$ and $t_3=\lambda_1.$ The hermitian metric
is then
\be
h=|z-\gamma_1|^2\bigg(1+\frac{\lambda_1^2}{|z-\gamma_1|^2}\bigg)
=t_3^2+(x_1-t_1)^2+(x_2-t_2^2),
\ee  
and the above formula, 
with normalization constant $K=3/(8\pi)$,
gives the moduli space metric
\be
g_{\mu\nu} dt_{\mu}dt_{\nu}=\frac{dt_1^2+dt_2^2+dt_3^2}{t_3^2}.
\ee
As advertised, 
this is indeed the metric of hyperbolic space, in upper half space 
coordinates. 

The moduli space of inversion symmetric hyperbolic 2-monopoles is 
four-dimensional and is obtained from the action of $SO(3)$ on the
one-parameter family of $D_2$ symmetric 2-monopoles described earlier.
It would be interesting to use the above approach to compute the metric
on this moduli space and to compare with both the Atiyah-Hitchin metric
for Euclidean 2-monopoles and Hitchin's metric, with $k=6$ in the
notation of \cite{Hi2}, which is an algebraic metric on the spectral
curves of precisely these hyperbolic 2-monopoles. 

As the moduli space metric is invariant under $SO(3)$ spatial rotations then
the one-parameter dihedral families discussed in the previous section, being 
obtained as fixed point sets of a finite subgroup of this $SO(3)$ action,
are automatically geodesics with respect to this metric. This relies on the
fact that, for the examples considered, there are no hyperbolic monopoles 
with the given symmetry and charge
that lie outside the JNR ansatz, which follows from known results
on the dimensions of spaces of symmetric Euclidean monopoles.   

\section{Conclusion}\quad\
For a specific relation between the curvature of hyperbolic space 
and the magnitude of the Higgs field at infinity, we have been able to 
obtain a complete description of a large class of hyperbolic $N$-monopoles.
We have presented 
simple explicit formulae for the spectral curve and the rational map, 
in terms of free data given by $N+1$ points on the
sphere together with positive real weights. 
This complements recent work that provided an explicit formula for 
the Higgs field in terms of the same data.
A number of symmetric examples
have been presented, including one-parameter families that are hyperbolic
analogues of geodesics that describe Euclidean monopole scattering.
We have derived an integral expression for an interesting 
 metric on the space of this
data and in future work we plan to investigate this aspect further.  

\appendix
\section{Appendix}\quad\
In this appendix we prove the rational map formula (\ref{JNRratmap})
directly using the definition
(\ref{ratmap}) together with the ADHM matrix (\ref{JNRM}) and the change of
basis matrices given by (\ref{vmatrix}) and (\ref{smatrix}).

The proof involves a formal expansion in $z^{-1}$.
We define the coefficients $q_I$ and $Q_I$ by
\be
L(z-M)^{-1}L^t=\sum_{I=1}^\infty q_Iz^{-I},
\label{exp1}
\ee
\be
\left\lbrace\sum^N_{i=0}\sum^N_{j=i+1}\lambda_i^2\lambda_j^2(\gamma_i-\gamma_j)^2\prod^N_{\substack{k=0\\k\neq i,j}}(z-\gamma_k)\right\rbrace/\left\lbrace\left(\sum^N_{i=0}\lambda_i^2\right)\left(\sum^N_{j=0}\lambda_j^2\prod^N_{\substack{k=0\\k\neq j}}(z-\gamma_k)\right)\right\rbrace
=\sum_{I=1}^\infty Q_Iz^{-I}.
\label{exp2}
\ee
The prove the rational map formula we need to 
show that $q_I=Q_I,$ which we accomplish by proving that both sets of 
coefficients satisfy the same inductive relation, together with $q_1=Q_1$.  

To begin, we expand the left hand side of (\ref{exp1}) to give
\be
q_I=LM^{I-1}L^t.
\ee
From the $(N+1)\times (N+1)$ matrix $S$, given by (\ref{smatrix}),
we define the $N\times (N+1)$ matrix $S^\prime$ by removing the top row of $S$.
Furthermore, we define this removed row to be 
$S^{\prime\prime}$. With this decomposition of $S$ and the corresponding 
decomposition (\ref{ADHMmatrix}) of the ADHM matrix, 
equation (\ref{JNRM}) becomes
\be
L=S^{\prime\prime} \Gamma V\quad\textrm{ and }\quad M=S^\prime \Gamma V,
\ee
and therefore
\be
q_{I+1}=LM^IL^t=S^{\prime\prime} \Gamma (VS^\prime \Gamma )^IVV^t \Gamma ^t(S^{\prime\prime})^t.
\ee
Using the fact that $S$ is an orthogonal matrix, it is easy to show that
\be
VS^\prime=\frac{1}{\lambda_0}(U^t- (S^{\prime\prime\prime})^tS^{\prime\prime})
\quad\textrm{ and }\quad 
VV^t=\frac{1}{\lambda_0^2}(1-(S^{\prime\prime\prime})^tS^{\prime\prime\prime}),
\ee
where 
$U$ is the $(N+1)\times N$ matrix obtained from the $N\times N$ 
identity matrix by adding an extra top row of zeros, and we have defined
the $N$-component row vector 
$S^{\prime\prime\prime}$ by $(S^{\prime\prime\prime})_i=(S^{\prime\prime})_{i+1}$.
Note that
\begin{align}
q_{I+1}&=
S^{\prime\prime} \Gamma \left(\frac{1}{\lambda_0}(U^t- (S^{\prime\prime\prime})^tS^{\prime\prime}) \Gamma \right)(VS^\prime \Gamma )^{I-1}VV^t \Gamma ^t(S^{\prime\prime})^t \nonumber \\
&=
\frac{1}{\lambda_0}S^{\prime\prime} \Gamma U^t \Gamma (VS^\prime \Gamma )^{I-1}VV^t \Gamma ^t(S^{\prime\prime})^t
-\frac{1}{\lambda_0}(S^{\prime\prime} \Gamma (S^{\prime\prime\prime})^t)q_I.
\end{align}
Continuing inductively, after some calculation, one finds that
\be
q_{I}=\sum^{I-1}_{J=1}a_Jq_{I-J}+b_I,
\label{qrel}
\ee
where we have introduced
\be 
a_I=-\frac{1}{\lambda_0^{I}}S^{\prime\prime} \Gamma (U^t \Gamma )^{I-1}(S^{\prime\prime\prime})^t
\quad\mbox{ and }\quad 
b_I=\frac{1}{\lambda_0^{I-1}}S^{\prime\prime} \Gamma (U^t \Gamma )^{I-1}VV^t \Gamma^t (S^{\prime\prime})^t.
\ee
  Since $U^t \Gamma $ is diagonal, $a_I$ and $b_I$ can be calculated to be 
\be
a_I=-p_N^2\sum^N_{j=1}\lambda_j^2\gamma_j^{I-1}(\gamma_j-\gamma_0),
\ee
\be
b_I=\left\lbrace\left(\sum^N_{j=1}
\lambda_j^2\gamma_j^{I-1}(\gamma_j-\gamma_0)^2\right)-p_N^2\left(\sum^N_{j=1}\lambda_j^2\gamma
_j^{I-1}(\gamma_j-\gamma_0)\right)\left(\sum^N_{k=1}\lambda_k^2(\gamma_k-
\gamma_0)\right)\right\rbrace p_N^2.
\ee
As the rational map is invariant under permutations of the poles and weights then the coefficients $q_I$ must be too. Taking (\ref{qrel}), exchanging
$(\gamma_0,\lambda_0)$ with $(\gamma_k,\lambda_k)$ and summing over $k$
from 0 to $N$  yields, after a long but
straightforward manipulation,
\begin{equation}\label{indrel}
(N+1)q_{I}=\sum^{I-1}_{J=1}\alpha_Jq_{I-J}+\beta_I,
\end{equation}
where
\be
\alpha_I=-p_N^2\sum^N_{j=0}\sum^N_{k=0}\lambda_j^2\gamma_j^{I-1}
\left(\gamma_j-\gamma_k \right),
\ee
and
\be
\beta_I=p_N^4\sum^N_{j=0}\sum^N_{k=j+1}\lambda_j^2\lambda_k^2(\gamma_j-\gamma_k)\left((N+1)(\gamma_j^{I}-\gamma_k^{I})-\left(\sum^N_{l=0}\gamma_l\right)(\gamma_j^{I-1}-\gamma_k^{I-1})
\right).
\ee

We now show that the induction relation \eqref{indrel} is also true for the $Q_I$. From (\ref{exp2}), after multiplying by the denominator of the
left hand side and cancelling an overall factor of 
$\prod^N_{j=0}(z-\gamma_j)$, we find that   
\be
p_N^2\sum^N_{i=0}\sum^N_{j=i+1}\lambda_i^2\lambda_j^2(\gamma_i-\gamma_j)\left(\frac{1}{z-\gamma_i}-\frac{1}{z-\gamma_j}\right)=\left(\sum^\infty_{I=1}{Q_I}{z^{-I}}\right)\left(\sum^N_{j=0}\frac{\lambda_j^2}{z-\gamma_j}\right).
\ee
Expanding this relation in $z^{-1}$ and comparing coefficients produces the 
induction relations
\begin{equation}\label{indrel2}
Q_{I-1}\tilde{\alpha}_1+Q_{I-2}\tilde{\alpha}_2+\dots+Q_1\tilde{\alpha}_{I-1}=\tilde{\beta}_{I}p_N^2,
\end{equation}
where
\be
\tilde{\alpha}_I=\sum^N_{j=0}\lambda_j^2\gamma^{I-1}_j
\quad \textrm{ and } \quad \tilde{\beta}_I=\sum^N_{j=0}\sum^N_{k=j+1}
\lambda_j^2\lambda_k^2(\gamma_j-\gamma_k)(\gamma_j^{I-1}-\gamma_k^{I-1}).
\ee
We will call \eqref{indrel2} the $I$-th of these induction relations.
The $\tilde{\alpha}_I$ are related to the $\alpha_I$ by
\be
(N+1)\tilde{\alpha}_{I+1}=\tilde{\alpha}_I\left(\sum^N_{k=0}\gamma_k\right)-\frac{\alpha_I}{p_N^2}.
\ee
Substituting this identity into \eqref{indrel2} gives
\be
\sum^{I-1}_{J=1}Q_{I-J}\left(p_N^2(N+1)\tilde{\alpha}_{J+1}+\alpha_J\right)=\left(\sum^N_{k=0}\gamma_k
\right)\tilde{\beta}_{I}p_N^4.
\ee
Subtracting this from $p_N^2(N+1)$ times the $(I+1)$-th  of the induction relations \eqref{indrel2}, 
gives
\be
(N+1)Q_{I}-\sum^{I-1}_{J=1}\alpha_JQ_{I-J}=p_N^4\left((N+1)\tilde{\beta}_{I+1}-\left(\sum^N_{k=0}\gamma_k\right)\tilde{\beta}_{I}\right)=\beta_I.
\ee
This shows that $Q_I$ and $q_I$ satisfy the same induction relation.

It is easy to check that
\be
q_1=Q_1=p_N^4\sum^N_{i=0}\sum^N_{j=i+1}\lambda_i^2\lambda_j^2(\gamma_i-\gamma_j)^2,
\ee
so $q_I=Q_I$ for all $I$, and this completes the proof.

\section*{Acknowledgements}
PMS thanks Nuno Rom\~ao for useful discussions.
This work is funded by the EPSRC grant EP/K003453/1 and 
the STFC grant ST/J000426/1.
AC acknowledges STFC for a PhD studentship.

\end{document}